\documentclass[article,nojss,shortnames]{jss}
\usepackage{thumbpdf,lmodern}
\usepackage{orcidlink}

\usepackage{amsmath}
\usepackage{amssymb}
\usepackage{mathtools}
\usepackage{multirow}
\usepackage{dsfont}
\usepackage{adjustbox}

\def\spacingset#1{\renewcommand{\baselinestretch}%
{#1}\small\normalsize} \spacingset{1}

\mathchardef\mhyphen="2D

\usepackage{booktabs}
\usepackage{bm}
\usepackage[inline]{enumitem}
\usepackage{rotating}

\author{Benjamin M\"uller~\orcidlink{0000-0003-3557-6926}\\University of Innsbruck
   \And Nikolaus Umlauf~\orcidlink{0000-0003-2160-9803}\\University of Innsbruck
   \And Johannes Seiler~\orcidlink{0000-0001-5714-9234}\\University of Innsbruck
   \AND Kenneth Harttgen~\orcidlink{0000-0002-8499-5281}\\ETH Zurich
   \And Stefan Lang~\orcidlink{0000-0003-0739-3858}\\University of Innsbruck
}
\Plainauthor{M\"uller~B., Umlauf~N., Seiler~J., Harttgen~K., Lang~S.}

\title{{\fontsize{16.5}{16.5}\selectfont Simultaneous Estimation and Model Choice for Big Discrete Time-to-Event Data with Additive Predictors}}
\Shorttitle{Estimation and Model Choice for Big Discrete Time-to-Event Data}

\Abstract{
Discrete-time hazard models are widely used when event times are measured in intervals or are not precisely observed. While these models can be estimated using standard generalized linear model techniques, they rely on extensive data augmentation, making estimation computationally demanding in high-dimensional settings. In this paper, we demonstrate how the recently proposed Batchwise Backfitting algorithm, a general framework for scalable estimation and variable selection in distributional regression, can be effectively extended to discrete hazard models. Using both simulated data and a large-scale application on infant mortality in sub-Saharan Africa, we show that the algorithm delivers accurate estimates, automatically selects relevant predictors, and scales efficiently to large data sets. The findings underscore the algorithm’s practical utility for analysing large-scale, complex survival data with high-dimensional covariates.

}

\Keywords{Discrete time-to-event analysis; Additive predictors; Big data; Batchwise backfitting; Automated variable selection}
\Address{
  Benjamin M\"uller, Nikolaus Umlauf, Johannes Seiler, and Stefan Lang\\
  \textit{Department of Statistics, University of Innsbruck, Innsbruck, Austria}\\
  E-mail: \email{benjamin.mueller@uibk.ac.at},\\
  \phantom{E-mail: }\email{nikolaus.umlauf@uibk.ac.at},\\
  \phantom{E-mail: }\email{johannes.seiler@uibk.ac.at},\\
  \phantom{E-mail: }\email{stefan.lang@uibk.ac.at}\\
   \\
  Kenneth Harttgen\\
  \textit{Development Economics Group, ETH Zurich, Zurich, Switzerland} and\\
  \textit{NADEL Center for Development and Cooperation, ETH Zurich, Zurich, Switzerland}\\
  E-mail: \email{kenneth.harttgen@nadel.ethz.ch}\\
}

\begin{document}
\sloppy

% \readme{Note: Statistical Modelling (SAGE) uses British English.}

% introduction --------------------
%% word count: xxx (Introduction to Conclusion)
\section{Introduction}\label{sec:intro}
Modelling the time until an event occurs is a central task in many areas of applied statistics. This field, commonly referred to as time-to-event or survival analysis, has a wide range of applications, from modelling survival outcomes of children to predicting the duration of marriages or the failure time of mechanical components \citep[e.g.,][]{burstein_mapping_2019, musick_change_2015, christodoulou_water_2011}. 

While much of the research in this area focuses on continuous-time models, discrete-time approaches are often more suitable when event times are measured in intervals, as is often the case in social science and health research (e.g., age of the children in months in child survival analyses). Foundational work by \cite{allison_discrete-time_1982} established the basis for discrete-time survival models, and subsequent developments have introduced extensions to enhance their flexibility \citep{fahrmeir_smoothing_1996}.

In parallel, the development of generalized additive models (GAMs) has advanced statistical modelling by allowing for nonlinear, smooth effects of covariates \citep{hastie_generalized_1986}. Building on this, innovations in smoothing techniques \citep{eilers_flexible_1996}, efficient estimation algorithms \citep{wood_thin_2003, wood_stable_2004, wood_generalized_2017}, and Bayesian extensions \citep{fahrmeir_bayesian_2001, brezger_generalized_2006} have made GAMs highly adaptable. 

These innovations have also influenced time-to-event analysis, leading to growing interest in flexible additive models for survival data \citep{tutz_flexible_2004, tutz_modeling_2016, berger_semiparametric_2018}. More recently, machine learning-inspired techniques based on tree methods have grown in popularity in this field \citep{schmid_survival_2016, puth_tree-based_2020, spuck_flexible_2023}.

Modern applications of time-to-event analysis increasingly involve large-scale and high-dimensional data sets, which present significant computational and statistical challenges (e.g., complex nonlinear effects and automated variable selection). In such contexts, efficient estimation methods and automatic variable selection are critical. However, to the best of our knowledge, scalable and interpretable approaches for additive discrete time-to-event models remain underdeveloped and insufficiently tested.

This paper addresses this gap by extending the recently proposed Batchwise Backfitting algorithm by \cite{umlauf_scalable_2024} to additive discrete time-to-event models. Our contribution lies in introducing a scalable estimation strategy that enables simultaneous model fitting and variable selection for big data sets. We evaluate the performance of the Batchwise Backfitting algorithm through a comprehensive simulation study and a real-world application.

The simulation study applies the method across a range of settings, including data sets with up to ten million observations and numerous informative and non-informative covariates. The results demonstrate that the proposed approach achieves comparable or superior performance to established alternatives in terms of estimation accuracy, variable selection, and estimation time. This suggests that the method is well-suited for large-scale, high-dimensional time-to-event data, providing researchers with a practical tool for tackling complex survival analyses that are challenging to address with existing approaches.

To demonstrate the practical utility of the method, we apply Batchwise Backfitting to model infant mortality in ten sub-Saharan African countries. The data set includes approximately $350000$ individual children and $26$ potential explanatory variables, representing a complex, high-dimensional setting. The results highlight the algorithm's ability to identify key determinants of child survival while maintaining scalability and interpretability.

The remainder of this paper is structured as follows: Section~\ref{sec:modelestimation} introduces the model and estimation framework, including the mathematical formulation of the model framework, as well as the additive predictors and details on the estimation approach. Section~\ref{sec:simulation} presents the simulation study, detailing the basic settings and their variations, as well as specifying the performance evaluation metrics. The results of the different methods regarding estimation accuracy and efficiency are discussed at the end of this section. Section~\ref{sec:application} applies Batchwise Backfitting to real-world data on infant mortality in sub-Saharan Africa, highlighting key findings, important factors and variable effects. Finally, Section~\ref{sec:conclusion} concludes with a discussion of the results, their implications, and potential directions for future research.

\pagebreak

% model and estimation --------------------
\section{Flexible discrete hazard models}
\label{sec:modelestimation}
\subsection{Model specification}\label{sec:model}
We consider a discrete time-to-event framework in which time is represented as a sequence of discrete periods, and each individual experiences at most one event. An event is defined as a one-time transition from one state to another (e.g., death, dropout, or system failure). While we use the term \emph{individual} for convenience, the concept can refer to any observational unit, including persons, households, or organisations. The model formulation and notation used in the following build on the frameworks of \citet{tutz_modeling_2016} and \citet{berger_semiparametric_2018}.

Although time is generally conceptualised as continuous, in many practical applications it is recorded or analysed in discrete intervals (e.g., days, months, or years) depending on the nature of data collection \citep{tutz_modeling_2016}. The exact timing of an event is often unknown and is instead observed only within a specific interval. Formally, let $t \in \{1, 2, \ldots, k\}$ index the discrete time intervals, which may either reflect naturally discrete time or grouped continuous time, with boundaries $[0, a_1), [a_1, a_2), \ldots, [a_{k-1}, \infty)$. For $n$ individuals, let $T_i \in \{1, \ldots, k\}$ denote the event time for individual~$i$, where $T_i = t$ implies that the event occurred in the interval $[a_{t-1}, a_t)$. The timing of the event is modelled in relation to $p$ explanatory variables $\mathbf{x}_i = (x_{i1}, \ldots, x_{ip})^\top$.

Censoring is a common feature in time-to-event data, arising when the event time for an individual is not fully observed. The most frequent case is right censoring, which occurs when the start of observation is known, but the event has not occurred by the end of the observation period. For right-censored data, the observed time is defined as $t_i := \min(T_i, C_i)$, where $T_i$ is the (possibly unobserved) event time and $C_i \in \{1, \ldots, k\}$ denotes the censoring time for individual~$i$. Other forms of censoring, such as left or interval censoring, are not considered in this work.

The hazard function is the central quantity in modelling time-to-event data. In a discrete-time setting, it is defined as the conditional probability that an event occurs at time~$t$, given that the individual has ``survived'' up to that time. For a given set of explanatory variables $\mathbf{x}_i$, $i = 1, \ldots, n$, the discrete-time hazard function is defined as
\begin{equation}\label{eqn:hazard}
\lambda(t \mid \mathbf{x}_i) = P(T_i = t \mid T_i \geq t, \mathbf{x}_i), \quad t = 1, \ldots, k,
\end{equation}
which represents the probability of transitioning from the initial to the terminal state during interval $t$, conditional on survival up to $t$ and covariates $\mathbf{x}_i$.

The corresponding survival function is given by
\begin{equation}\label{eqn:survival}
S(t \mid \mathbf{x}_i) = P(T_i > t \mid \mathbf{x}_i) = \prod_{r=1}^{t} \left(1 - \lambda(r \mid \mathbf{x}_i)\right), \quad t = 1, \ldots, k,
\end{equation}
and denotes the probability that the event occurs after time~$t$, or equivalently, that the individual survives the interval $[a_{t-1}, a_t)$.

A general model for the discrete hazard function in Equation~\eqref{eqn:hazard} can be defined as
\begin{equation}\label{eqn:hazardmodel}
\lambda(t \mid \mathbf{x}_i) = h(\eta_{it}),
\end{equation}
where $h(\cdot)$ is a strictly increasing link function, such as the logit function $h(x) = \exp(x)/(1 + \exp(x))$, and $\eta_{it}$ is the structured additive predictor for individual~$i$ at time~$t$. In a flexible specification, $\eta_{it}$ takes the form \citep{fahrmeir_regression_2022}
\begin{equation}\label{eqn:addpredictor}
\eta_{it} = f_0(t; \boldsymbol{\beta}_0) + f_1(\mathbf{x}_{i}; \boldsymbol{\beta}_1) + \ldots + f_p(\mathbf{x}_{i}; \boldsymbol{\beta}_p),
\end{equation}
where $f_0(\cdot)$ models the baseline hazard over time, and $f_1(\cdot), \ldots, f_p(\cdot)$ represent (possibly nonlinear) covariate effects. 

Each function $f_j(\cdot)$ is approximated using basis functions\:\textendash\:such as penalized splines \citep[e.g.,][]{eilers_flexible_1996, wood_thin_2003}\:\textendash\:with the (potentially nonlinear) effects expressed as
\begin{equation*}
\mathbf{f}_j(\mathbf{X}_j; \boldsymbol{\beta}_j) = \mathbf{X}_j \boldsymbol{\beta}_j, \quad j = 0, \ldots, p,
\end{equation*}
where $\mathbf{X}_j$ denotes the design matrix and $\boldsymbol{\beta}_j$ the associated coefficients. Stacking all components, the predictor of the additive discrete hazard model can be expressed in matrix form as
\begin{equation*}
\boldsymbol{\eta} = \sum_{j=0}^{p} \mathbf{X}_j \boldsymbol{\beta}_j,
\end{equation*}
where $\boldsymbol{\eta}$ is the vector over all individuals and risk intervals. This corresponds to a flexible predictor commonly used in structured additive regression (STAR; \citealp{fahrmeir_penalized_2004, brezger_generalized_2006, lang_multilevel_2014}).

\subsection{Penalized likelihood}\label{sec:loglik}
In discrete hazard models, the primary focus is on estimating the model specified in equations~\eqref{eqn:hazardmodel} and~\eqref{eqn:addpredictor}. If the censoring mechanism is assumed to be independent of the explanatory variables related to the event time (i.e., uninformative censoring), the likelihood function simplifies considerably, as censoring does not affect the distribution of the event time. The individual contribution to the likelihood is given by
\begin{equation*}
L_i = c_i \, P(T_i = t_i)^{\delta_i} \, P(T_i > t_i)^{1 - \delta_i},
\end{equation*}
where the indicator variable $\delta_i = \mathbb{I}(T_i \le C_i)$ distinguishes between observations with observed events ($\delta_i = 1$) and right-censored observations ($\delta_i = 0$), and $c_i := P(C_i \geq t_i)^{\delta_i} \, P(C_i = t_i)^{1 - \delta_i}$ accounts for the censoring mechanism. Under the assumption of uninformative censoring, $c_i$ can be treated as a constant and is thus omitted from the likelihood used to estimate $\boldsymbol{\beta} = (\boldsymbol{\beta}_0, \ldots, \boldsymbol{\beta}_p)^\top$.

The likelihood can be expressed as a sum of individual contributions, where each contribution corresponds to the log-likelihood of a Bernoulli-distributed random variable. Specifically, for each individual~$i$, we define a binary indicator~$y_{is}$ for each discrete time interval~$[a_{s-1}, a_s)$ up to the observed or censored time~$t_i$, with $y_{is} = 1$ if the event occurs in interval~$s$, and $y_{is} = 0$ otherwise. Since each individual can experience the event at most once, the data can be represented as a sequence of independent Bernoulli trials, each with success probability equal to the discrete hazard $\lambda(t \mid \mathbf{x}_i)$. The total log-likelihood of the discrete hazard model then becomes
\begin{align}\label{eqn:loglik-eta}
\ell(\boldsymbol{\beta}; \mathbf{y}, \mathbf{X}) 
&\propto \sum_{i=1}^{n} \sum_{r=1}^{t_i} \left[ y_{ir} \log \lambda(r \mid \mathbf{x}_i) + (1 - y_{ir}) \log \left(1 - \lambda(r \mid \mathbf{x}_i) \right) \right] \\
&= \sum_{i=1}^{n} \sum_{r=1}^{t_i} \left[ y_{ir} \log h(\eta_{ir}) + (1 - y_{ir}) \log \left(1 - h(\eta_{ir}) \right) \right], \nonumber
\end{align}
where $\lambda(r \mid \mathbf{x}i) = h(\eta_{ir})$ denotes the discrete hazard at time~$r$ for individual~$i$, with $h(\cdot)$ representing a suitable strictly increasing link function, and $\eta_{ir}$ the structured additive predictor as defined in Equation~\eqref{eqn:addpredictor}. The vector $\mathbf{x}_i = (x_{i1}, \ldots, x_{ip})^\top$ contains the covariate values for individual~$i$, and the full design matrix is given by $\mathbf{X} = (\mathbf{X}_0, \ldots, \mathbf{X}_p)$, where each $\mathbf{X}_j$ corresponds to the basis functions used to represent the model term $f_j(\cdot)$, and $\mathbf{X}_0$ represents the design matrix for the smooth baseline effect $f_0(\cdot)$ in Equation~\eqref{eqn:addpredictor}.

To prevent the model terms $f_j(\cdot)$ from overfitting, regularization is introduced through a penalized log-likelihood of the form
\begin{equation*}
\ell(\boldsymbol{\beta}, \boldsymbol{\tau}; \mathbf{y}, \mathbf{X}) \propto \ell(\boldsymbol{\beta}; \mathbf{y}, \mathbf{X}) - \sum_{j=0}^{p} \boldsymbol{\beta}_j^\top \mathbf{P}_j(\boldsymbol{\tau}_j) \boldsymbol{\beta}_j,
\end{equation*}
where $\boldsymbol{\tau}_j$ is a (possibly vector-valued) smoothing parameter that controls the degree of regularization applied to each model component $f_j(\cdot)$, and $\mathbf{P}_j(\boldsymbol{\tau}_j)$ is a corresponding quadratic penalty matrix determined by the structural assumptions on $f_j(\cdot)$.

For example, in the case of univariate smooth terms modelled via P-splines, $\boldsymbol{\tau}_j$ reduces to a scalar $\tau_j$, and the penalty matrix typically takes the form $\mathbf{P}_j(\tau_j) = \tau_j \mathbf{K}_j$, where $\mathbf{K}_j$ is based on second-order difference penalties on $\boldsymbol{\beta}_j$ (\citealp{fahrmeir_regression_2022}, Section~8.1.2, pp. 449-464). The smoothing parameters $\boldsymbol{\tau}_j$ thus govern the overall functional form of the estimated effects.

\subsection{Estimation}\label{sec:estimation}
Estimation of discrete hazard models reduces to the estimation of a binary regression model but requires prior data augmentation. This involves expanding the original data set by adding pseudo-observations for each individual~$i$ and each time point $t = 1, \ldots, t_i$ during which the individual is at risk. 

This data augmentation step is best illustrated with an example. Imagine we are interested in $t = 1, 2, 3$ and the initial data set in Table~\ref{tab:augment} on the left.

\begin{table}[htb!]
\centering
\begin{tabular}{c @{\hspace{0.5cm}} | @{\hspace{0.5cm}} c}
\multicolumn{1}{l}{\textit{\hspace{0.5cm}Before augmentation:}} & \multicolumn{1}{l}{\textit{\hspace{0.3cm}After augmentation:}} \\
% left table
\begin{tabular}[t]{cccccc}
\textit{id} & \textit{t} & \textit{y} & \textit{x}\textsubscript{1} & \textit{x}\textsubscript{2} & \textit{x}\textsubscript{3}\\\hline
1 & 1 & 0 & x\textsubscript{11} & x\textsubscript{12} & x\textsubscript{13}\\
2 & 3 & 1 & x\textsubscript{21} & x\textsubscript{22} & x\textsubscript{23}\\
3 & 2 & 1 & x\textsubscript{31} & x\textsubscript{32} & x\textsubscript{33}\\
4 & 5 & 0 & x\textsubscript{41} & x\textsubscript{42} & x\textsubscript{43}\\
\end{tabular}
&
% right table
\begin{tabular}[t]{cccccc}
\textit{id} & \textit{t} & \textit{y} & \textit{x}\textsubscript{1} & \textit{x}\textsubscript{2} & \textit{x}\textsubscript{3}\\\hline
1 & 1 & 0 & x\textsubscript{11} & x\textsubscript{12} & x\textsubscript{13}\\
2 & 1 & 0 & x\textsubscript{21} & x\textsubscript{22} & x\textsubscript{23}\\
2 & 2 & 0 & x\textsubscript{21} & x\textsubscript{22} & x\textsubscript{23}\\
2 & 3 & 1 & x\textsubscript{21} & x\textsubscript{22} & x\textsubscript{23}\\
3 & 1 & 0 & x\textsubscript{31} & x\textsubscript{32} & x\textsubscript{33}\\
3 & 2 & 1 & x\textsubscript{31} & x\textsubscript{32} & x\textsubscript{33}\\
4 & 1 & 0 & x\textsubscript{41} & x\textsubscript{42} & x\textsubscript{43}\\
4 & 2 & 0 & x\textsubscript{41} & x\textsubscript{42} & x\textsubscript{43}\\
4 & 3 & 0 & x\textsubscript{41} & x\textsubscript{42} & x\textsubscript{43}\\
\end{tabular}
\end{tabular}
\caption{\label{tab:augment} Toy example for the data augmentation step: Initial data set before augmentation on the left and data set after augmentation on the right.}
\end{table}

The individual identification variable is labelled as $\textit{id}$, $\textit{t}$ is the time variable, $\textit{y}$ is the event indicator, where $1$ indicates an event and $0$ otherwise, and $\textit{x}_1,~\textit{x}_2$ and $\textit{x}_3$ are the time-invariant explanatory variables. Note that the individuals $1$ and $4$ are right-censored, as we do not observe the time to the event, since no event occurred in the observation period ($\textit{y} = 0$). 

Following augmentation, the data set takes the form shown on the right in Table~\ref{tab:augment}. In this example, a total of five rows are added, two for individuals $2$ and $4$ and one for individual $3$. The time variable $t$ needs to cover the individual risk period, the event indicator $y$ is $0$ for the added rows, and the individual explanatory variables $\textit{x}_1,~\textit{x}_2$ and $\textit{x}_3$ are duplicated. 

In many applications, the resulting data expansion can be substantial, often leading to very large data sets that exceed the capacity of standard estimation algorithms. Efficient estimation methods for the GAM framework are therefore essential for handling such high-dimensional settings (see, e.g., \citealp{wood_generalized_2017, lang_multilevel_2014}). Despite this, model term selection for such large models is difficult and not really considered in these implementations.

Recently, \citet{umlauf_scalable_2024} proposed the Batchwise Backfitting algorithm\:\textendash\:an efficient approach for scalable estimation of distributional regression models with automatic selection of model terms. Since the discrete hazard model as stated in Equation~\eqref{eqn:hazard} is a special case of a distributional regression model, the resulting Newton--Raphson-type updating equations can be written in an iterative form. These equations are used to maximize the log-likelihood in Equation~\eqref{eqn:loglik-eta} and estimate the coefficient vectors $\boldsymbol{\beta_j}$. For iteration $l+1$, the update for component $j$ is given by
\begin{equation}\label{eqn:bfit}
  \boldsymbol{\beta}_{j}^{[l+1]} =
    (\mathbf{X}_{j}^\top\mathbf{W}\mathbf{X}_{j} +
      \mathbf{P}_{j}(\boldsymbol{\tau}_{j}))^{-1}\mathbf{X}_{j}^\top\mathbf{W}(
      \mathbf{z} - \boldsymbol{\eta}_{-j}^{[l+1]}),
\end{equation}
where $\mathbf{z} = \boldsymbol{\eta}^{[l]} + \mathbf{W}^{-1}\mathbf{u}$ is a vector of working observations and $\mathbf{u} = \partial \ell(\boldsymbol{\beta}; \mathbf{y}, \mathbf{X}) / \partial \boldsymbol{\eta}$ is the score vector. $\mathbf{W} = -\mathrm{diag}(\partial^2 \ell(\boldsymbol{\beta}; \mathbf{y}, \mathbf{X}) / \partial \boldsymbol{\eta}^2)$ is a \emph{diagonal weight matrix of size $N \times N$}, where $N = \sum_{i=1}^n t_i$ is the total number of augmented binary observations. Each diagonal entry in $\mathbf{W}$ is evaluated at the current state $\boldsymbol{\beta}^{[l]}$. The expression in Equation~\eqref{eqn:bfit} corresponds to a backfitting update for the model term $f_j(\cdot)$, with $\boldsymbol{\eta}_{-j}$ denoting the predictor excluding the $j$-th model component.

The backfitting loop in Equation~\eqref{eqn:bfit} cycles through model terms $j = 0, \ldots, p$, and is iterated until convergence, e.g., when the relative change in the coefficients falls below a prespecified threshold. Smoothing parameters $\boldsymbol{\tau}_j$ can be estimated using a stepwise selection procedures (similar to \citealp{belitz_simultaneous_2008}), optimizing each component sequentially within adaptive search intervals and using information criteria (e.g., AIC or BIC). As mentioned above, in many practical cases, $\boldsymbol{\tau}_j$ reduces to a scalar. For further algorithmic details, see \citet{umlauf_bamlss_2018}.

Scalable estimation is achieved by using only a randomly selected batch of the data. In discrete hazard models, where each individual contributes $t_i$ observations (one for each time interval during which they are at risk) to the augmented data set, the natural sampling unit is the individual. If an individual is selected for a batch, then all the corresponding observations are included. Formally, let
\begin{equation*}
\mathbf{i} = \left(1,\ \ldots,\ \sum_{i=1}^{n} t_i \right)^\top = \left( \mathbf{i}_1^\top,\ \ldots,\ \mathbf{i}_n^\top \right)^\top
\end{equation*}
denote the vector of stacked row indices in the augmented data set, where
\begin{equation*}
\mathbf{i}_i = \left( \sum_{m=1}^{i-1} t_m + 1,\ \ldots,\ \sum_{m=1}^{i} t_m \right)^\top
\end{equation*}
gives the indices corresponding to individual $i$, for $i = 1, \ldots, n$. A batch is then defined by selecting a random subset $ s\subseteq \{1, \ldots, n\}$ of individuals and using all corresponding observations, i.e., $\mathbf{i}_s = \bigcup_{i \in s} \mathbf{i}_i$. 

The corresponding response vector $ \mathbf{y}_{[\mathbf{i}_s]} $ and covariate matrix $ \mathbf{X}_{[\mathbf{i}_s]} $ are used to compute a stochastic updating step of the form
\begin{eqnarray} \label{eqn:bbfit}
\boldsymbol{\beta}_{j}^{[l+1]} &=& (1 - \nu) \cdot \boldsymbol{\beta}_{j}^{[l]} + \\
&& \hspace*{-1.7cm}\nu \cdot (\mathbf{X}_{[\mathbf{i}_s], j}^\top\mathbf{W}_{[\mathbf{i}_s]}\mathbf{X}_{[\mathbf{i}_s], j} +  \mathbf{P}_{j}(\boldsymbol{\tau}_{j}))^{-1}\mathbf{X}_{[\mathbf{i}_s], j}^\top\mathbf{W}_{[\mathbf{i}_s]}(\mathbf{z}_{[\mathbf{i}_s]} - \boldsymbol{\eta}_{[\mathbf{i}_s], -j}^{[l+1]}) \nonumber \\
&=& (1 - \nu) \cdot \boldsymbol{\beta}_{j}^{[l]} + \nu \cdot \boldsymbol{\beta}_{[\mathbf{i}_s], j},
\nonumber
\end{eqnarray}
where $\nu$ is the step length control parameter (or \textit{learning rate}) specifying the amount of which $\boldsymbol{\beta}_{j}^{[l]}$ is updated in the direction of the new estimate $\boldsymbol{\beta}_{[\mathbf{i}_s], j}$ on batch $[\mathbf{i}_s]$. Note that the working weights $\mathbf{W}_{[\mathbf{i}_s]}$ and the score vectors $\mathbf{u}$ used to compute $\mathbf{z}_{[\mathbf{i}_s]}$ are evaluated on the $\left[l\right]$-th estimate $\boldsymbol{\beta}_{j}^{[l]}$. In each iteration, Equation~\eqref{eqn:bbfit} is evaluated on exactly one batch $[\mathbf{i}_s]$, such that the computational burden can be reduced considerably. 

The updating function defined in Equation~\eqref{eqn:bbfit} can be applied in several ways. For instance, if the step length control parameter is set to a small value, such as $\nu = 0.1$, and only the best-fitting model term $f_j(\cdot)$ is updated in each iteration, the algorithm mimics a boosting-type approach. Specifically, the decision to update a model term is based on its log-likelihood contribution, evaluated on another random batch $[\tilde{\mathbf{i}}_s]$. This design helps mitigate overfitting, as demonstrated by \citet{umlauf_scalable_2024}. Similarly, smoothing parameters $\boldsymbol{\tau}_j$ are selected by minimizing an information criterion (e.g., AIC) on a different independent batch. This simultaneous selection of model terms and smoothing parameters offers a major advantage as it eliminates the need to determine an optimal stopping iteration, which is typically required in classical boosting algorithms (e.g., via computationally intensive cross-validation).

The algorithm is considered to have converged when the ``out-of-sample'' log-likelihood evaluated on other batches no longer improves. To better account for uncertainty, a common strategy is to refit the selected model with $\nu = 1$, mimicking a resampling step. Estimates are then based on the last iterations, like using Markov chain Monte Carlo (MCMC) simulation. This algorithm has proven to have excellent model term selection performance and can be applied to very large data sets, as it is often the case in discrete hazard models. For a detailed description of the algorithm, please refer to \citet{umlauf_scalable_2024}.

The number of batches and thus the number of iterations $L$ should be chosen sufficiently large such that no (major) further improvements in the out-of-sample log likelihood can be observed. In this work, $L$ is between $200$ and $500$ in the simulation study and $1000$ in the application. Furthermore, a reasonable batch size $M$ must be defined, which in previous exercises was chosen between $10000$ and $50000$ (e.g. \cite{umlauf_scalable_2024} or \cite{seiler_high-resolution_2025}). Depending on the complexity and type of data, the batches can be smaller or must be larger to ensure sufficient coverage of the information in the data. For both the simulation and the application, we use a batch size of $M = 20000$.

% simulation --------------------
\section{Simulation study}\label{sec:simulation}
A comprehensive simulation study is conducted to investigate the performance of Batchwise Backfitting for estimating discrete time-to-event models, especially in the context of extensive data sets. We apply Batchwise Backfitting and suitable benchmark methods in different settings and evaluate estimation accuracy, variable selection performance and estimation time. To ensure robustness of the results, each setting is replicated $250$ times for each method under study.

\subsection{Design and basic setting}\label{sec:sim:designbasic}
Using the framework developed in Section~\ref{sec:modelestimation}, data is simulated for individuals $i = 1, \ldots, n$ and time $t = 1, \ldots, k$. In the basic setting, the number of individuals $n$ is set to $5000$. The number of time points $k$ is kept at $20$ in all settings of the simulation study. 

The predictor is defined as
\begin{equation}\label{eqn:simpred}
    \eta_{it} = f_{0}(t) + f_{1}(x_{i1}) + \ldots + f_{9}(x_{i9}),
\end{equation}
where $f_0(t)$ represents the baseline hazard function, $f_1(x_{i1}), \ldots, f_4(x_{i4})$ are the effects of informative variables $x_{i1}, \ldots, x_{i4}$, and $f_5(x_{i5}) \equiv 0, \ldots, f_9(x_{i9}) \equiv 0$ are the effects of non-informative/noise variables $x_{i5}, \ldots, x_{i9}$.

\begin{figure}[htb!]
    \centering
    \includegraphics[width=0.5\textwidth]{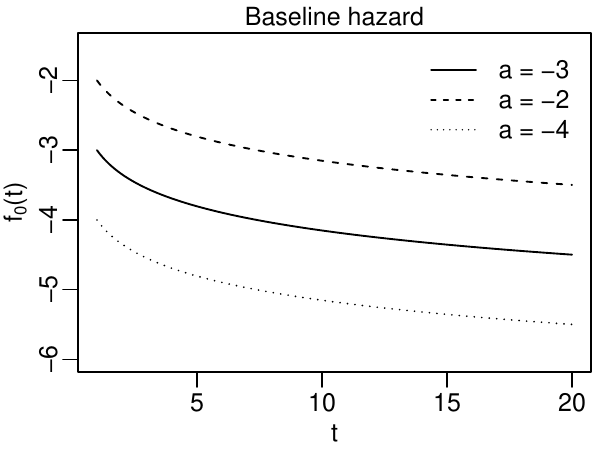}
    \vspace{-0.3cm}\caption{\label{fig:sim:basehazs} Specifications of the baseline hazard function $f_0(t)$. $a = -3$ is used in the basic setting. $a = -2$ and $a = -4$ are variations to investigate the influence of different event frequencies.}
\end{figure}

The \textit{baseline hazard} function $f_0(t)$ is specified decreasing logarithmically over time as
\begin{equation*}
    f_0(t) = a - \frac{1}{2} \log(t).
\end{equation*}
In the basic setting $a = -3$, which corresponds to an event frequency of approximately $10\%$. This specification is shown in Figure~\ref{fig:sim:basehazs} as solid curve.

\begin{figure}[htb!]
    \centering
    \includegraphics[width=0.9\textwidth]{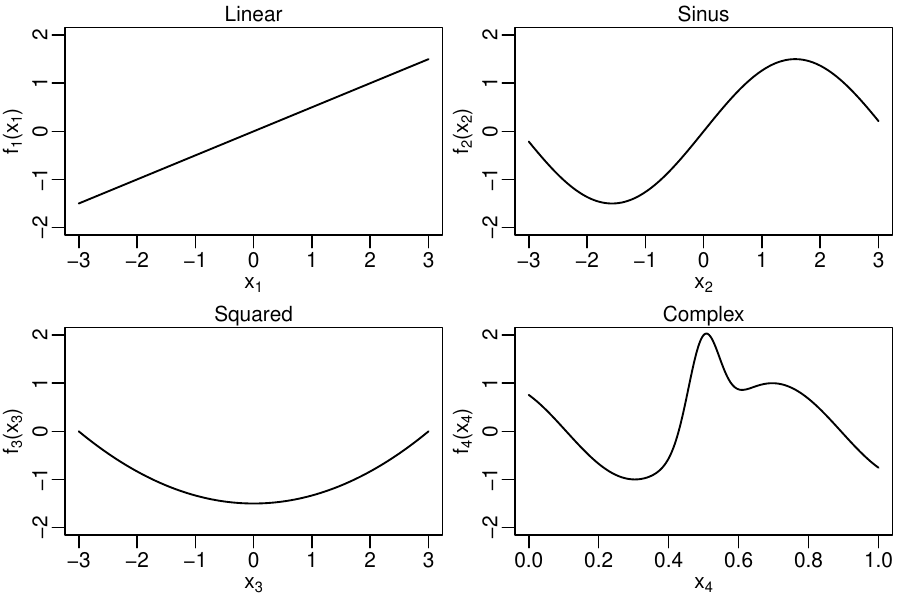}
    \vspace{-0.3cm}\caption{\label{fig:sim:eff} Univariate informative effects on $\eta_{it}$.}
\end{figure}

Equidistant design points are generated from the interval $[0, 1]$ for the variable $x_{i4}$ and from the interval $[-3, 3]$ for the remaining variables. The effects of the informative variables are defined as
\begin{align*}
            f_1(x_1) &= 0.5 \cdot x_1,           &   f_2(x_2) &= 1.5 \cdot sin(x_2),\\
            f_3(x_3) &= x_3^{\frac{2}{6}} - 1.5, &   f_4(x_4) &= sin(2 \cdot (4 \cdot x_4 - 2)) + 2 \cdot e^{-(16^2) \cdot (x_4-0.5)^2},
\end{align*}
see Figure~\ref{fig:sim:eff}. The unique values of the effects $f_1(x_{1}), \ldots, f_4(x_{4})$ are scaled to have a specific standard deviation ($SD$), which is set to $SD = 1$ in the basic setting. In the following, we refer to $f_1(x_1)$ as \textit{linear}, $f_2(x_2)$ as \textit{sinus}, $f_3(x_3)$ as \textit{squared}, $f_4(x_4)$ as \textit{complex} and $f_5(x_5), \ldots, f_9(x_9)$ as \textit{noise 1} to \textit{noise 5}. 

The event indicator $y_{it}$ is derived by comparing $\lambda(t\vert\mathbf{x}_i)$ with random draws from the uniform distribution $u_{it} \sim \mathcal{U}(0,~1)$, i.e.,
\begin{align*}
y_{it} = 
\begin{cases} 
1, & \text{if } u_{it} \le \lambda(t \vert \mathbf{x}_i), \\ 
0, & \text{otherwise}.
\end{cases}
\end{align*}
For each individual, only the first occurrence of an event (i.e., the first time $y_{it} = 1$) is considered as the event time $T_i$, and a random censoring time $C_i$ is generated, where $C_i \sim \mathcal{U}(\{1, 2, \dots, 20\})$. The individual's observation time is then defined as $t_i = \min(T_i,~C_i,~20)$. This results in approximately $(n \cdot k)/2$ rows in the augmented data set used for estimation. In the basic setting (where $n = 5000$), the augmented data set consists of around $50000$ rows.

\subsection{Further settings}
The further settings are variations of the basic setting to evaluate the performance of the estimation methods under different circumstances. In each setting, one feature is varied, while all other components remain as in the basic setting. The number of individuals, the baseline hazard specification and the scaling of the effects are varied and a predictor with a spatial effect is simulated:

\begin{itemize}
    \item \textbf{Number of individuals} \bm{$n$}: \sloppy To investigate the performance for both small and large data sets, simulations are performed with number of individuals $n = 1000; 10000; 50000; 100000; 500000$ and $1000000$. This results in augmented data sets that range from approximately $10000$ to $10000000$ rows. 
    
    \item \textbf{Baseline hazard specification} \bm{$f_0(t)$}: Alternative baseline hazard specifications are considered to analyse the influence of different event frequencies. Simulations are performed with $a = -2$ and $a = -4$, which corresponds to an event frequency of approximately $20\%$ and $5\%$, respectively. The two variations are also visualised in Figure~\ref{fig:sim:basehazs}.
    
    \item \textbf{Effect scaling} \bm{$SD$}: The scaling of the unique values of the effects is varied to $SD = 0.5$ and $SD = 2$ to investigate the influence of different effect sizes. However, since the scaling of the effects does not reveal any remarkable pattern, we refrain from presenting and discussing these settings in the remainder. 

    \item \textbf{Spatial effect} \bm{$f_{spa}(lon,lat)$}: In addition to the univariate effects, a bivariate spatial effect $f_{spa}(lon,~lat)$ is added to the model. It is defined as 
    \begin{equation*}
        f_{\text{spa}}(lon,~lat) = 2.5 \cdot \sin(lon) \cdot \sin\left(0.5 \cdot lat\right) - 0.3,
    \end{equation*}
    and the updated predictor is $\theta_{it} = \eta_{it} + f_{spa}(lon,~lat)$, where $\eta_{it}$ is defined in Equation~\eqref{eqn:simpred}. For $lon$ and $lat$ equidistant design points are sampled from the interval $[-3, 3]$. The spatial effect is visualised in Figure~\ref{fig:sim:spatial}.
    \begin{figure}[htb!]
        \centering
        \includegraphics[width=0.38\textwidth]{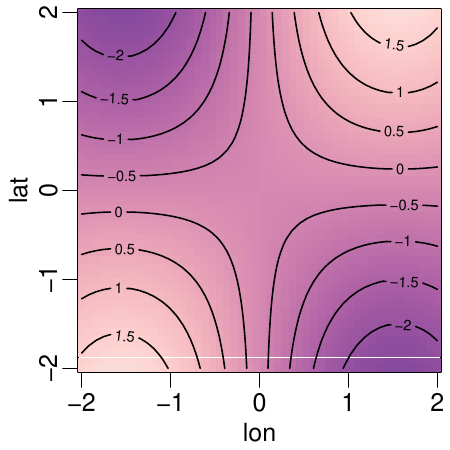}
        \vspace{-0.3cm}\caption{\label{fig:sim:spatial} Bivariate spatial effect on $\theta_{it}$.}
    \end{figure}
\end{itemize}

\subsection{Methods and implementation}
In the following, we refer to Batchwise Backfitting as BBFIT. The method is implemented in the \proglang{R} package \pkg{bamlss} \citep{umlauf_bamlss_2018, umlauf_bamlss_2021} and the Batchwise Backfitting algorithm \citep{umlauf_scalable_2024} can be used by setting \code{optimizer = opt\_bbfit} in the main function \code{bamlss()}. The method is applied as outlined in Section \ref{sec:modelestimation}, see also \cite{umlauf_scalable_2024} or \cite{seiler_high-resolution_2025} for a less technical description.

Model comparison is based on the out-of-sample AIC. We use a combination of the boosting and resampling variants of the algorithm in a two-step procedure. In the first step, the boosting variant is executed for $L = 200$ iterations including all possible covariates with the default step length $\nu = 0.1$. To perform a variable selection, the argument \code{select = TRUE} is specified. This step serves as a covariate preselection for the second step, the resampling variant of the algorithm. The second step is also executed for $L = 200$ iterations, with the first $100$ iterations discarded as a burn-in phase. In this step, the model contains only the selected covariates. The standard step length $\nu = 1$ is applied, and slice sampling \citep{neal_slice_2003} is used for the smoothing parameters.

The batches are generated by sampling individuals at random and including all observations of a sampled individual in a batch to account for the augmented data structure (see Section~\ref{sec:estimation}). We define a (maximum) batch size $M = 20000$. This implies that for the setting with $1000$ individuals, corresponding to an augmented data set with around $10000$ rows, each batch consists of the entire model data set including all individuals. For settings with more individuals, only a portion of the data set is included in each batch until $M$ is reached.

The model specification follows the approach introduced by \citet{wood_thin_2003}, using thin-plate regression splines \code{s()} for the baseline hazard function and all univariate effects of the explanatory variables, with the default basis dimension (\code{k = 10}). For the bivariate spatial effect, a full tensor product smooth is used in its default form via \code{te()}, as described in \citet{wood_lowrank_2006}.

The performance of BBFIT is compared with two benchmark methods:

\begin{enumerate}
    \item As a basic benchmark, we use an estimation method for generalized linear models with stepwise model selection based on the Akaike information criterion (AIC). Specifically, the function \code{stepAIC()} from the \proglang{R} package \pkg{MASS} \citep{venables_modern_2002} with the specification \code{direction = ‘both’} is applied. The baseline hazard function and all effects of the explanatory variables are modelled with third-degree polynomials. In the following, we refer to this approach as GLM.
    \item For more flexible specifications, an estimation method for generalised additive models optimised for (very) large data sets is employed. Here we use the \code{bam()} function \citep{wood_generalized_2015, wood_generalized_2017} from the \proglang{R} package \pkg{mgcv} \citep{wood_thin_2003, wood_stable_2004, wood_fast_2011, wood_smoothing_2016, wood_generalized_2017}. The smoothing parameter is estimated with a fast version of a restricted maximum likelihood approach. This is the default approach of \code{bam()} specified by \code{method = "fREML"}. For variable selection, we set \code{select = TRUE}. In the following, this approach is referred to as BAM.
\end{enumerate}

\subsection{Performance measures} 
The measures aim to evaluate the performance of the estimation methods in terms of effect estimation, variable selection and estimation time: 

\begin{itemize}
    \item \textbf{Effect estimation}: To evaluate the performance of the methods in terms of effect estimation, all estimated effects are plotted against the true effect and the means and quantiles are displayed to detect systematic biases. Furthermore, the MSE of the estimated effects is calculated, defined as the mean of the squared differences between the true effects and the estimated effects, i.e. $MSE_{f} = \frac{1}{n} \sum_{i=1}^{n} (\hat{f}(x_i) - f(x_i))^2$, where $f(\cdot)$ is the true, $\hat{f}(\cdot)$ is the estimated effect and $x_1, \ldots, x_n$ are design points spanning the entire range of $x$.
    
    \item \textbf{Variable selection}: For the selection performance, we examine selection frequencies, with particular attention to the noise/uninformative variables. This frequency is computed by dividing the number of times each variable is selected by the total number of replications. For the methods GLM and BBFIT, selection is determined by whether a variable is included or excluded from the model. For BAM, however, there is no such clear rule, since effects of uninformative variables are estimated to be close to zero, but never exactly zero. To establish a selection rule for BAM, we evaluate several thresholds based on \textit{equivalent degrees of freedom (EDF)} and \textit{p-values}, and identify \textit{p-value $< 0.01$} as the best criterion, see Section \ref{subsec:results} below for more details. 

    \item \textbf{Estimation time}: Estimation time refers to the time required to estimate the model, which is exclusively the time that the functions need for the estimation. The estimation time is based on replications that are run on a ``standard'' PC (see Section~\ref{sec:simulation:computationaldetails}) to ensure practical relevance.
\end{itemize}

\subsection{Computational details}\label{sec:simulation:computationaldetails}
The simulation was run on the high-performance cluster (HPC) infrastructure LEO4 of the University of Innsbruck. This HPC infrastructure runs on a Linux system (CentOS $7$), and $50$ computing nodes with Intel Xeon (Broadwell/Skylake) processors with up to $3000$ gigabyte (GB) available memory.
To analyse the estimation time for the methods under study, some replications were re-run on a ``standard'' PC, as this is a typical setup for users. This PC runs with the Ubuntu $24.04.1$ LTS operating system. The model is a Lenovo ThinkPad T14s Gen $3$ with an AMD Ryzen $7$ PRO $6850$U processor, AMD Radeon Graphics (x $16$) and $32$ GB of available memory.

\subsection{Results}\label{subsec:results}
\subsubsection{Basic setting}
\begin{figure}[htb!]
    \centering
    \includegraphics[width=0.85\textwidth]{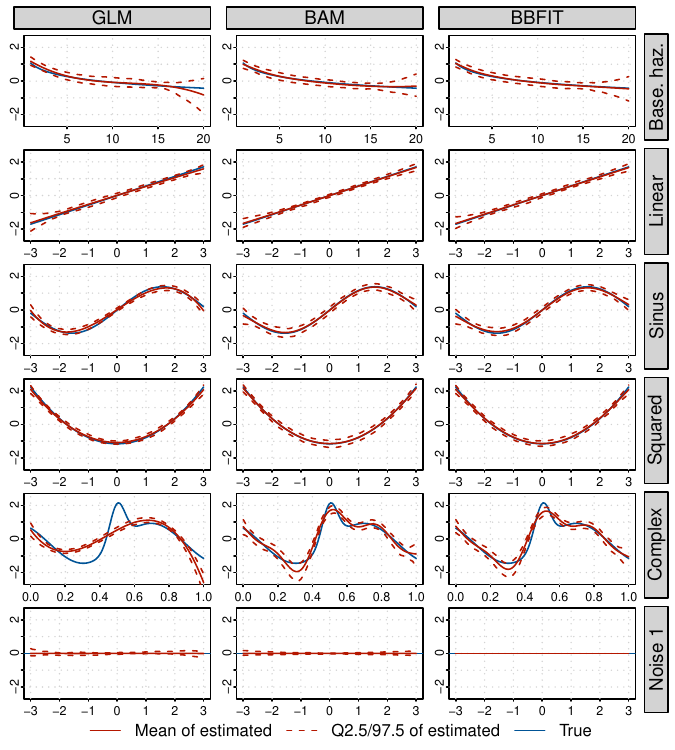}
    \vspace{-0.3cm}\caption{\label{fig:sim:basicesti} Estimated effects of all informative variables and the first noise variable in the basic setting (see Section \ref{sec:sim:designbasic}).}
\end{figure}

\begin{figure}[htb!]
    \centering
    \includegraphics[width=0.82\textwidth]{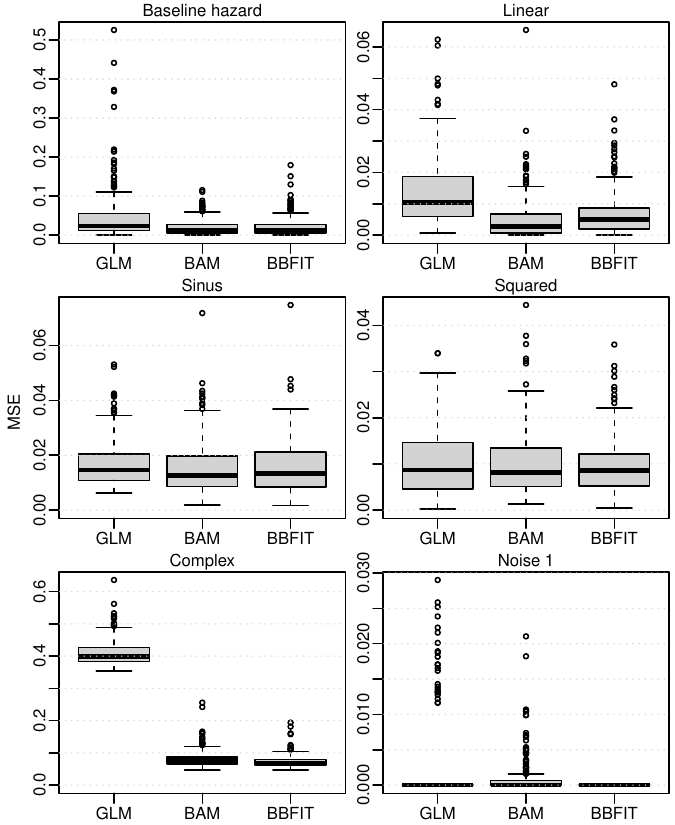}
    \vspace{-0.3cm}\caption{\label{fig:sim:basicrmse} MSE of the baseline hazard, the complex effect and the first noise effect in the basic setting (see Section \ref{sec:sim:designbasic}). Note: The MSEs of BBFIT for Noise 1 (bottom right) are always zero as the noise variables are never selected.}
\end{figure}

Figure \ref{fig:sim:basicesti} shows the estimated effects of all informative variables and the first noise variable, while Figure~\ref{fig:sim:basicrmse} shows the corresponding distribution of the MSE. Since the other noise variables show very similar patterns, their results are omitted here, but can be found in Figures~\ref{fig:apx:basicestinoise} and~\ref{fig:apx:basicrmsenoise} in the Appendix \ref{sec:apx:sim:basic}.

To analyse the estimated effects in more detail, we created a web app that allows replications and effects to be visualised separately or as an overlay in a dynamic plot. The app also offers the possibility to explore the further settings described below. It can be accessed via \href{https://bmueller5000.github.io/bb4sa-shinylive/}{bmueller5000.github.io/bb4sa-shinylive/} and the corresponding data can be found in \href{https://fileshare.uibk.ac.at/d/d78d077c7d184beab2ba/}{fileshare.uibk.ac.at/d/d78d077c7d184beab2ba/}.

\begin{figure}[htb!]
    \centering
    \includegraphics[width=0.75\textwidth]{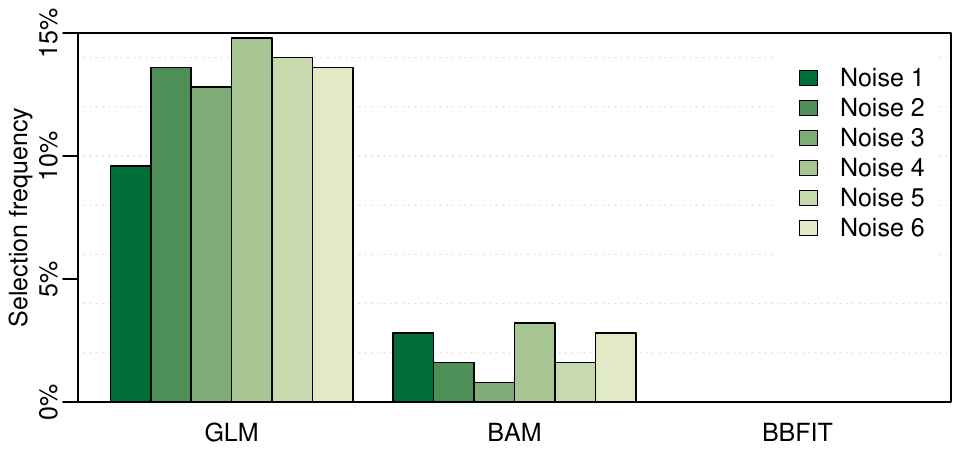}
    \vspace{-0.3cm}\caption{\label{fig:sim:basicsel} Selection frequencies of the noise variables in the basic setting (see Section \ref{sec:sim:designbasic}).}
\end{figure}

\begin{figure}[htb!]
    \centering
    \includegraphics[width=0.75\textwidth]{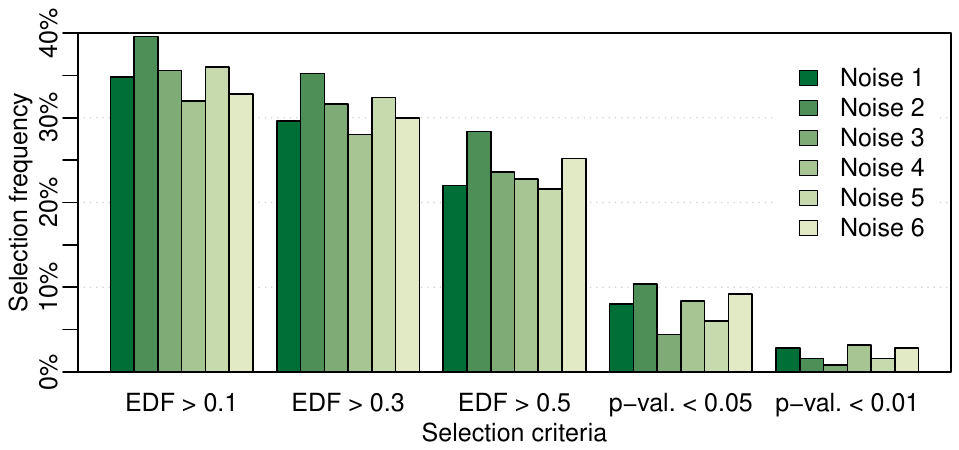}
    \vspace{-0.3cm}\caption{\label{fig:sim:basicselmgcvcrit} Selection frequencies of the noise variables for different selection criteria of BAM in the basic setting (see Section \ref{sec:sim:designbasic}).}
\end{figure}

Further results regarding the selection frequencies are discussed in Figure~\ref{fig:sim:basicsel}. Since the informative variables are always selected, we restrict the presentation to the noise variables. As mentioned above, the selection approach of BAM does not provide a clear selection rule. Therefore, different thresholds for the \textit{EDF} and \textit{p-values} are tested as selection criteria and visualised in  Figure~\ref{fig:sim:basicselmgcvcrit}.

From Figures~\ref{fig:sim:basicesti}~to~\ref{fig:sim:basicselmgcvcrit} and the web app, we draw the following conclusions:

\begin{itemize}
\item {\em Bias:} All methods provide largely unbiased estimates for the linear, sinus, and the squared effect $f_1,\:f_2,\:f_3$ and substantial bias for the complex effect~$f_4$. This bias is, as expected, more pronounced for the GLM method, since polynomials of degree three are not flexible enough to capture such complex effects. The GLM method also shows a clear bias for the baseline hazard~$f_0$. BAM shows a small bias in the baseline hazard~$f_0$ at higher $t$ values, which is almost invisible in Figure~\ref{fig:sim:basicesti}. Surprisingly, this bias increases with the number of individuals, see the more detailed discussion below. 

\item {\em MSE:} The main findings from the visible inspection of the bias are reflected in the MSEs. For the linear, sinus, and the squared effect $f_1,\:f_2,\:f_3$ the MESs are almost zero for all methods. For the baseline hazard~$f_0$, BAM and BBFIT yield lower MSEs than GLM and for the complex effect~$f_4$, GLM is clearly outperformed. Regarding the noise variables, BBFIT achieves MSEs of exactly zero in all replications. BAM shows low but non-zero MSE values due to occasional false positive selections, while GLM produces considerably higher MSEs for Noise 1.

\item {\em Selection frequencies:} Compared to the \textit{p-values}, selection via the \textit{EDF} consistently leads to higher selection frequencies for the noise variables in BAM (Figure~\ref{fig:sim:basicselmgcvcrit}). The best results are obtained with the rule \textit{p-value} $< 0.01$ which is therefore used for BAM in the remainder. Figure~\ref{fig:sim:basicsel} shows exceptional selection performance of BBFIT as the noise variables are never selected. The selection frequency of BAM (based on the rule \textit{p-value $< 0.01$}) varies between $0.8\%$ and $3.2\%$ and GLM is not competitive with selection frequencies between $10\%$ and $15\%$. 
\end{itemize}

\subsubsection{Further settings: Number of individuals}
\begin{figure}[htb!]
    \centering
    \includegraphics[width=\textwidth]{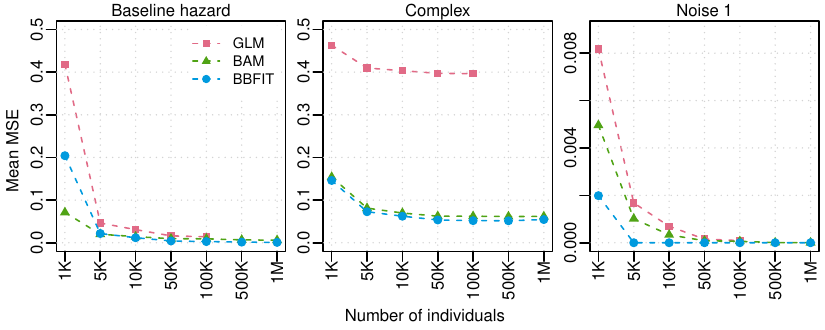}
    \vspace{-0.3cm}\caption{\label{fig:sim:furthrmse} Mean MSE for different number of individuals of the baseline hazard, the complex effect and the first noise effect.}
\end{figure}

Figure \ref{fig:sim:furthrmse} shows the average MSE across all $250$ replications for varying numbers of individuals, focusing on the baseline hazard $f_0$, the complex effect $f_4$, and the first noise effect $f_5$. These three effects show the most notable patterns. A plot including all effects is provided in Appendix \ref{sec:apx:sim:furth}, Figure~\ref{fig:apx:furthrmseall}. Estimation using the GLM method was not feasible for data sets with more than $100000$ individuals and is therefore excluded from the analysis of large data sets.

With increasing numbers of individuals, the estimation accuracy of the methods increases, which is reflected in a decreasing average MSE. However, the average MSE is almost constant for all methods and effects from $5000$ individuals. As already discussed in the basic setting, GLM performs worst in the estimation of the complex effect $f_4$ due to model specification limitations. BAM and BBFIT perform more or less equally well for $50000$ to $1000000$ individuals, and for fewer individuals no method is superior to the other.

\begin{figure}[htb!]
    \centering
    \includegraphics[width=0.75\textwidth]{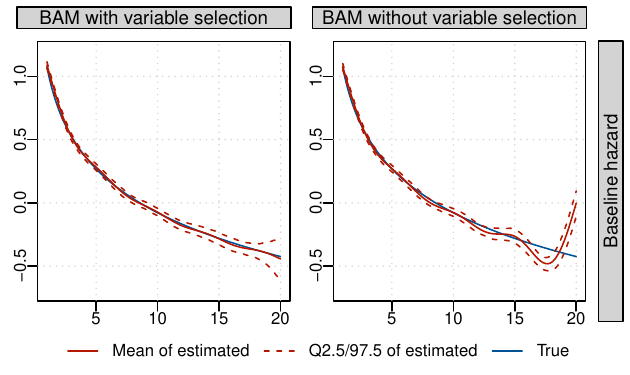}
    \vspace{-0.3cm}\caption{\label{fig:sim:furth500kindbl} Estimated effects of the baseline hazard with $500000$ individuals.}
\end{figure}

Figure \ref{fig:sim:furth500kindbl} displays the estimated baseline hazards based on $500000$ individuals, comparing the BAM method with variable selection (\code{select=TRUE}), which is used as the standard throughout this work, to the \code{bam()} method without variable selection (\code{select=FALSE}).

As mentioned above and clearly present with $500000$ individuals, BAM shows a bias in the baseline hazard at higher $t$ values. To investigate this bias further, we tested different methods for estimating the smoothing parameters, as well as varied the number of nodes and the smoothing basis for the splines. These variations did not resolve the issue. However, we found that the bias is not present when using the methods without variable selection, indicating that the variable selection approach might cause this issue. Further research is needed to test for systematic biases with the selection method of \code{bam()} in a discrete time-to-event data structure and in general, less special, data structures.

\begin{figure}[htb!]
    \centering
    \includegraphics[width=0.55\textwidth]{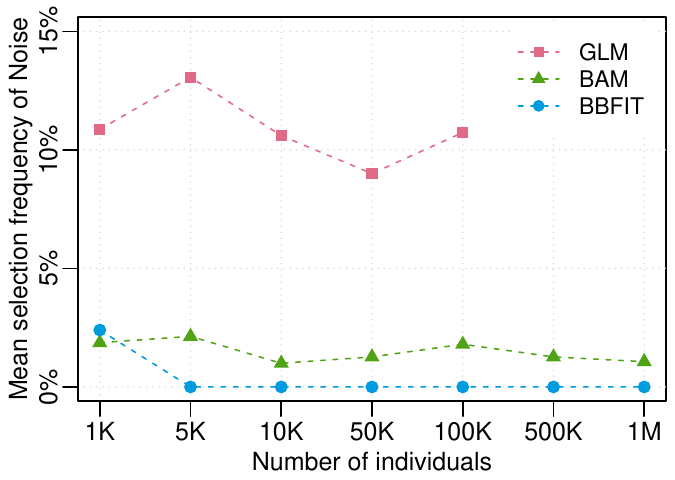}
    \vspace{-0.3cm}\caption{\label{fig:sim:furthselfreq} Mean selection frequency of the six noise variables for different number of individuals.}
\end{figure}

Figure \ref{fig:sim:furthselfreq} shows the mean selection frequency of the six noise variables for different number of individuals. The observed patterns are consistent for the individual noise variables, implying that the mean selection frequency effectively summarises the qualitative implications.

Independent of the number of individuals, the method GLM performs worst with a mean selection frequency of around $10\%$ to $12\%$. BAM performs significantly better with a selection rate around $1\%$ to $4\%$. The best selection performance is achieved using BBFIT with perfect selection for $5000$ or more individuals. Interestingly, the selection performance of all methods is relatively unaffected by the number of individuals. 

\begin{figure}[htb!]
    \centering
    \includegraphics[width=0.55\textwidth]{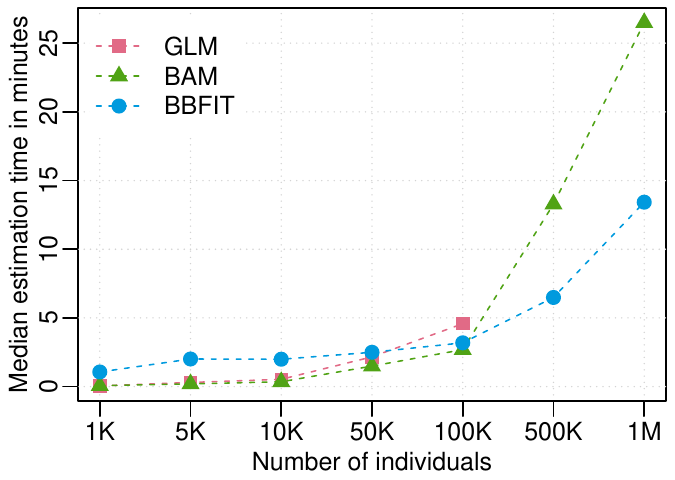}
    \vspace{-0.3cm}\caption{\label{fig:sim:furthestitime} Median estimation time in minutes for different number of individuals based on 25 replications run on a ``standard'' PC (see Section~ \ref{sec:simulation:computationaldetails}).}
\end{figure}

Figure \ref{fig:sim:furthestitime} shows the median estimation time for different numbers of individuals based on $25$ replications on a standard PC (see Section~ \ref{sec:simulation:computationaldetails}). The median estimation time for individuals up to $100000$ ranges from a few seconds to around $3$ minutes and is comparable for the three methods. For very large sample sizes of $500000$ or even $1000000$ individuals, BBFIT clearly outperforms BAM with median estimation time of BBFIT being about half as long as with BAM.

\begin{figure}[htb!]
    \centering
    \includegraphics[width=0.95\textwidth]{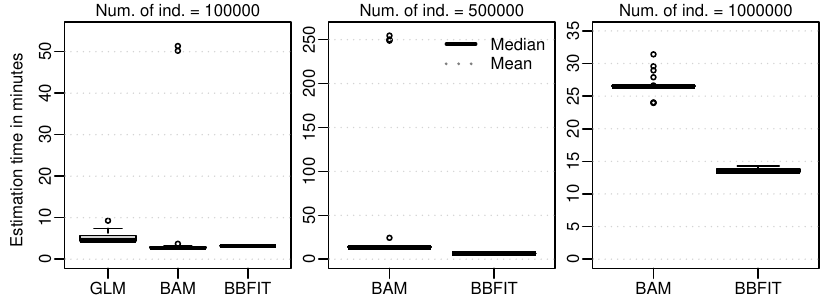}
    \vspace{-0.3cm}\caption{\label{fig:sim:furthestitimedist} Boxplots of estimation time in minutes for $100000$, $500000$ and $1000000$ individuals based on 25 replications run on a ``standard'' PC (see Section~ \ref{sec:simulation:computationaldetails}).}
\end{figure}

Figure \ref{fig:sim:furthestitimedist} further shows the distribution of estimation times for $100000$, $500000$ and $1000000$ individuals as box plots including the mean. While the estimation times for GLM and BBFIT are relatively stable, BAM shows some large outliers. For $500000$ individuals, for example, $4$ of $25$ replications have estimation times of about $250$ minutes, while the rest have about $13$ minutes, resulting in a large discrepancy between the mean and the median. These large outliers in estimation times are not only observed on our PC, but also occur when running the models on the HPC. For these exceptionally long estimation times, a warning is issued that the algorithm did not converge. However, when looking at the estimated effects of these replications, we found no qualitative differences compared to the replications with normal estimation times and no warning.

\subsubsection{Further settings: Baseline hazard specification}
\begin{figure}[htb!]
    \centering
    \includegraphics[width=0.55\textwidth]{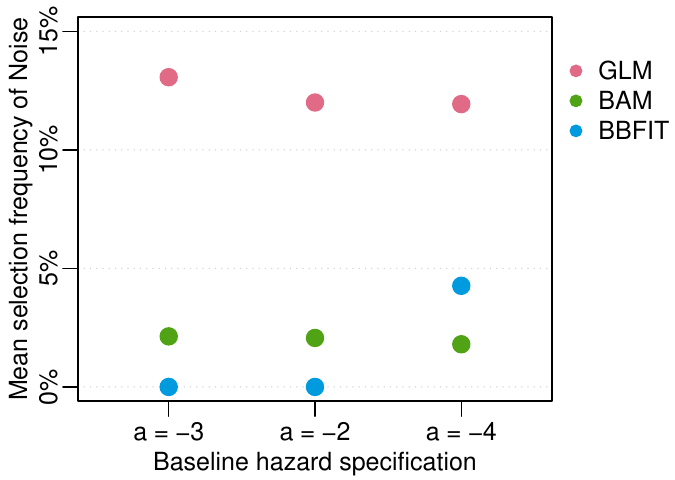}
    \vspace{-0.3cm}\caption{\label{fig:sim:furthselfreqbali} Mean selection frequency of the six noise variables for different baseline hazard specifications.}
\end{figure}

The alternative baseline hazard specifications $a = -2$ and $a = -4$ (see Figure~\ref{fig:sim:basehazs}) have no considerable influence on the estimation performance in terms of estimation accuracy of the methods under study. However, we find an effect on the selection frequencies. Figure~\ref{fig:sim:furthselfreqbali} shows the mean selection frequencies of the six noise variables for the three baseline hazard specifications. A notable result occurs with the specification $a = -4$ (event frequency of around $5\%$). Here, the selection performance of BBFIT rises from $0$ to about $5\%$ and is outperformed by BAM.

\subsubsection{Further settings: Spatial effect}
\begin{figure}[htb!]
    \centering
    \includegraphics[width=0.45\textwidth]{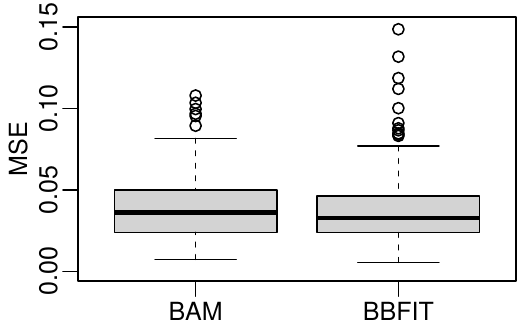}
    \vspace{-0.3cm}\caption{\label{fig:sim:furthspamse} MSE of bivariate spatial effect.}
\end{figure}

The bivariate spatial effect can only be added with BAM and BBFIT. The GLM method does not support any multivariate functional forms. Figure~\ref{fig:sim:furthspamse} shows the distributions of the MSE based on $250$ replications.

Both methods estimate the spatial effect equally well. BAM shows a marginally larger bias in some areas, but overall the MSE is similar for the two methods.

% application --------------------
\section{Application}\label{sec:application}
Despite considerable improvements in the past, infant mortality rates (i.e., the number of deaths of children under one year of age per $1000$ live births in a given year) remain high in some parts of the world. This is especially true in sub-Saharan Africa, where the most recently reported infant mortality rate was $4.4\%$ (i.e., $44$ per $1000$ live births) in $2023$ \citep{united_nations_childrens_fund_united_2025}. While this reflects a substantial reduction from $10.2\%$ in $1990$, the rate remains high compared to other regions in the world. For instance, the reported infant mortality rate in South Asia\:\textendash\:the second most affected region\:\textendash\:was $3\%$, and in North America, it was about ten times lower than in sub-Saharan Africa in the same year. In absolute terms, around $1.8$ million children under one year of age died in sub-Sahara Africa, corresponding to around $51\%$ of all infant deaths globally in $2023$.

Child mortality (including neonatal, infant, and under-five mortality) is a complex issue influenced by a range of factors, such as inadequate access to health care, poor nutrition, and environmental conditions. Therefore, child mortality rates serve as crucial indicators of overall human development. This is prominently recognized in the third Sustainable Development Goal (\emph{SDG~3})\:\textendash\:\emph{Good health and well-being} \citep{united_nations_sustainable_2015}. Promoting child health is a high priority in \emph{SDG~3}, with one of the targets being to reduce neonatal and under-five mortality rates in all countries by $2030$.

The analysis of infant mortality involves modelling the probability of death between birth and the first birthday \citep{Rutstein2006GuideDHS}. In our application, we employ a time-to-event model with an additive predictor to model and estimate infant mortality in ten countries in eastern sub-Saharan Africa.

\subsection{Data} \label{sec:app:data}
For modelling infant mortality in eastern sub-Saharan Africa, Demographic and Health Surveys (DHS, \cite{ICF:2004-2017}) provide representative and standardised survey data commonly used to monitor health related outcomes in low- and middle-income countries. This data is merged with remotely sensed data on climate, demography, environmental information to create a unique data source. The geo-spatial data was initially compiled for \cite{seiler_high-resolution_2025} to estimate the prevalence of anemia in children. They provide a detailed overview of the data and details of the preprocessing steps in the Section~\textit{Data} and especially in the \textit{Supplementary Information} \citep{seiler_supplementary_2025}.

In this application, we use a subset of that data on individual children from the eastern sub-Saharan Africa countries (based on the UN geoscheme classification, \citealp{united_nations_statistics_division_standard_2025}) Burundi, Ethiopia, Kenya, Malawi, Mozambique, Rwanda, United Republic of Tanzania, Uganda, Zambia, and Zimbabwe. Further data cleaning involved the exclusion of surveys older than the year $2000$, observations with missing covariates and observations born more than five years before the survey date. The resulting data set contains $351705$ individual children, of which about $4.5\%$ died in the first year of life. As discussed in Section~\ref{sec:modelestimation}, the data must be augmented before fitting the models. This augmentation step substantially increases the data set up to around $3.7$ million rows, which are ultimately used for estimation. 

All basic information for modelling mortality is recorded for these children. This includes an indicator whether the child is still alive at the time of observation (\texttt{dead}), as well as information on the age at the time of observation or the age at death on a monthly basis (\texttt{age}). In addition, the data set contains numerous potential explanatory variables, such as the age of the mother at birth (\texttt{magebirth}), information on the material wealth of the household (\texttt{ai}) and the average temperature anomalies at the household's location (\texttt{t2m2}). We can draw on a total of $25$ potential explanatory variables, which are listed in Table~\ref{tab:app:vars}. 

\begin{sidewaystable}
\centering
\caption{\label{tab:app:vars} Covariates included in the full model as used in the selection step.}
\begin{adjustbox}{width=0.8\textwidth}
\centering
\begin{small}
\begin{tabular}{p{2cm}p{2.5cm}p{2cm}p{5cm}p{7cm}}
& & & & \\
\textbf{Variable} & \textbf{Unit} & \textbf{Type} & \textbf{Description} & \textbf{Source/Reference} \\
\toprule
\multicolumn{5}{l}{\textit{Response}} \\
$\text{\code{dead}}$ & 1 if "dead"; 0 if "alive" & Binary & Death/living status at the day of the interview & \href{https://dhsprogram.com/}{DHS}; \cite{ICF:2004-2017} \\
\midrule
\multicolumn{5}{l}{\textit{Continuous or quasi-continuous covariates}}\\
$\text{\code{age}}$ & Months & Metric & Age of child & \href{https://dhsprogram.com/}{DHS}; \cite{ICF:2004-2017} \\
$\text{\code{ai}}$ & Index & Continuous & Asset index of the household & \href{https://dhsprogram.com/}{DHS}; \cite{ICF:2004-2017} \\
$\text{\code{altitude}}$ & Meters & Continuous & Altitude in m above sea level & \href{https://www.ncei.noaa.gov/access/metadata/landing-page/bin/iso?id=gov.noaa.ngdc.mgg.dem:316}{NOAA ETOPO}; \cite{AmanteETOPOelevation2009,ETOPOelevation2009} \\
$\text{\code{bord}}$ & Count & Metric & Birth order within household & \href{https://dhsprogram.com/}{DHS}; \cite{ICF:2004-2017} \\
$\text{\code{l\_distance}}$ & Kilometers & Continuous & Log of distance to the closest body of water & \href{https://gis.ess.washington.edu/}{GRG Washington}; \cite{nasajpl:2013waterbodies} \\
$\text{\code{hhs}}$ & Count & Metric & Household size & \href{https://dhsprogram.com/}{DHS}; \cite{ICF:2004-2017} \\
$\text{\code{higheduyear}}$ & Years & Metric & Highest completed year of schooling & \href{https://dhsprogram.com/}{DHS}; \cite{ICF:2004-2017} \\
$\text{\code{magebirth}}$ & Years & Metric & Age of mother at birth & \href{https://dhsprogram.com/}{DHS}; \cite{ICF:2004-2017} \\
$\text{\code{minc12}}$ & Prevalence & Continuous & Malaria incidence & \href{https://malariaatlas.org/}{Malaria Atlas Project}; \cite{weiss2019,BattleGethingPv2019} \\
$\text{\code{ndvi12}}$ & Index & Continuous & Normalised difference vegetation index & \href{https://climatedataguide.ucar.edu/}{GIMMS}; \href{https://lpdaac.usgs.gov/}{MODIS}; \cite{Tucker+Pinzon+ElSaleous:2005,Pinzon+Tucker:2014,Didan:2016} \\
$\text{\code{pre12}}$ & Meters & Continuous & Precipitation & \href{https://www.ecmwf.int/}{ERA5}; \cite{Hersbach+Bell+Thepaut:2020} \\
$\text{\code{rgdp}}$ & US\$ & Continuous & Real GDP of the country & \href{https://worldbank.org/}{WDI}; \cite{wb2022WDI} \\
$\text{\code{t2m12}}$ & Kelvin & Continuous & 2~m surface temperature & \href{https://www.ecmwf.int/}{ERA5}; \cite{Hersbach+Bell+Thepaut:2020} \\
$\text{\code{l\_ttcity}}$ & Hours & Continuous & Log of travel time to city & \href{https://malariaatlas.org/}{Malaria Atlas Project}; \cite{Weiss2018} \\
$\text{\code{l\_ttmotor}}$ & Hours & Continuous & Log of travel time to healthcare facility by motorised vehicle & \href{https://malariaatlas.org/}{Malaria Atlas Project}; \cite{Weiss2020travel} \\
$\text{\code{l\_ttwalk}}$ & Hours & Continuous & Log of travel time to healthcare facility by foot & \href{https://malariaatlas.org/}{Malaria Atlas Project}; \cite{Weiss2020travel} \\
$\text{\code{lon}; \code{lat}}$ & Degree & Continuous & Longitude and latitude coordinates & \href{https://dhsprogram.com/}{DHS}; \cite{ICF:2004-2017} \\
\multicolumn{5}{l}{\textit{Discrete covariates}} \\
$\text{\code{gender}}$ & "female"; "male" & Binary & Sex of the child & \href{https://dhsprogram.com/}{DHS}; \cite{ICF:2004-2017} \\
$\text{\code{d\_conf25;}}$\newline$\text{\code{d\_conf50;}}$\newline$\text{\code{d\_conf100}}$ & "yes" if $ x \geq 5$;\newline "no" if $ x < 5$ & Binary & Indicator of reported conflicts in the past within a buffer of $25/50/100~km$ & \href{https://ucdp.uu.se/}{UCDP}; \cite{Davies+Petterson+Odberg2022ucdp,sundberg2013} \\
$\text{\code{d\_nl20\_12}}$ & "yes" if $ x \geq 5$;\newline "no" if $ x < 5$ & Binary & Indicator of observed night-time light digital number values & \href{https://www.nature.com/articles/s41597-020-0510-y}{Night-time light}; \cite{LiYuyuMin2020nightlight} \\
$\text{\code{s\_lc12}}$ & Classification & Categorical & Land-cover classification & \href{https://lpdaac.usgs.gov/}{MODIS Land Cover}; \cite{FriedlHuang2010lc,FriedlSulla2015LC} \\
$\text{\code{s\_soil}}$ & Classification & Categorical & Soil type classification & \href{https://soilgrids.org/} {Soilgrids}; \cite{deSousa+Rossiter:2020soil} \\
\bottomrule
\end{tabular}
\end{small}
\end{adjustbox}
\end{sidewaystable}

\sloppy{These variables can be broadly categorised into child-related, maternal, socio-economic, community-related and climatic factors, all of which are considered potentially important explanatory factors associated with infant mortality.} The selection of variables is informed by the \textit{UNICEF Conceptual Framework on the Determinants of Maternal and Child Nutrition} \citep{unicef_unicef_2020}. While the framework focuses on undernutrition, it is particularly relevant for studying child survival outcomes, as nearly half of all child deaths can be attributed to some form of malnutrition \citep{black_maternal_2008}.

Previous research has identified several socio-demographic key factors that influence infant and child mortality. These include the sex of the child \citep[see e.g.,][]{Seiler2019u5m, Kiross2021MortalityETH, Shobiye2022MortalityNigeria}, which has been associated with survival in various contexts, and birth order \citep[see e.g.,][]{Hobcraft1985im}, which potentially affects health outcomes through parental investment and sibling competition. Maternal age at birth is also critical, with both younger and older mothers facing higher risks due to biological and socioeconomic factors \citep[see e.g.,][]{Seiler2019u5m, Shobiye2022MortalityNigeria}. Parental education, especially of the mother, can be considered to serve as a protective factor by improving health knowledge, child care practices, and health care utilization \citep[see e.g.,][]{Hammer2003icm, Ekholuenetale2020MortalityAfrica, Shobiye2022MortalityNigeria}. In addition, urban or rural residence was found to influence access to health care and living conditions, while household wealth likely influences nutrition, access to health care, and general living standards, all of which are closely linked to child mortality risk \citep[see e.g.,][]{Seiler2019u5m, Ekholuenetale2020MortalityAfrica}.

In addition to socio-demographic factors, the presence of armed conflict, along with environmental and climatic conditions, can provide valuable insights into explaining child mortality. For instance, children growing up near areas affected by armed conflict face a substantially higher risk of dying before their first and fifth birthdays \citep{wagner2018conflict-mortality, boerma2019conflict-mortality}. Furthermore, malaria, one of the leading infectious diseases in tropical regions, is strongly associated with both infant and child mortality \citep[see e.g.,][]{BattleGethingPv2019, weiss2019}. Malaria disproportionately affects young children, particularly in regions with high transmission rates, leading to an increased risk of death from severe complications of the disease. Moreover, improved modelling techniques and the growing availability of socio-economic, environmental, and climatic geo-spatial data have proven invaluable in explaining infant and child mortality. For example, data on factors such as accessibility, land cover, precipitation, surface temperature, and vegetation are particularly effective when high-resolution geo-spatial modelling is applied \citep[see e.g.,][]{golding_mapping_2017, burstein_mapping_2019}. These factors, including climatic and environmental conditions, can act as proxies for infectious diseases or indicators of development, making them essential for assessing child and infant survival. 

Before presenting the full modelling framework and the results, we examine the covariate \textit{age of the mother at birth} (\texttt{magebirth}) in more detail, which is found to be one of the most important predictors in our analysis (see Figure~\ref{fig:app:upfreqscontrib}). This variable is used to illustrate how individual covariates can be explored and interpreted within our modelling approach. Specifically, we use it to discuss key analytical steps, including the structure of the survival data and the presentation of time-to-event outcomes. Starting with basic descriptive patterns in the data, we show in Section~\ref{sec:app:results} how the estimated effects can be visualised and interpreted. This illustration is intended to make the full results easier to follow and to highlight the value of variable-specific analyses.

\begin{figure}[htb!]
    \centering
    \includegraphics[width=\textwidth]{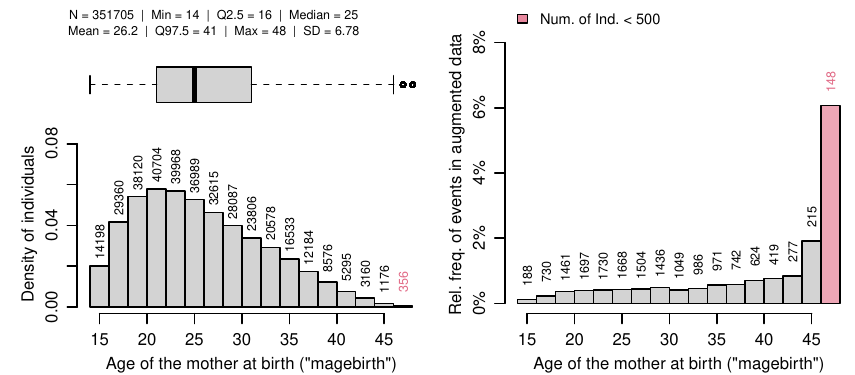}
    \vspace{-0.3cm}\caption{\label{fig:app:estivarsmagebirthdes} Distribution of the covariate \textit{age of the mother at birth} (\texttt{magebirth}) on the left and relative frequency of events (deaths) in the augmented data on the right.}
\end{figure}

Figure \ref{fig:app:estivarsmagebirthdes} shows the distribution of \texttt{magebirth} on the left, where age groups with fewer than $500$ observations are highlighted in red.

Based on our sample, most mothers give birth between the ages of $18$ and $28$, with a median age of $25$. The number of observations declines steadily after age $22$, with only $356$ mothers aged $46$ or older at the time of birth. On the right, the relative frequency of observed events (deaths) is displayed, which shows a contrasting pattern: relatively low mortality rates below age $44$, followed by a pronounced increase above that age, and a particularly high rate in the oldest age group. These insights are useful, when interpreting the estimated effects and other survival-specific outcomes in Section~\ref{sec:app:results} below.

\subsection{Model} \label{sec:app:model}
The probability of death in month $t = 1, \dots 11$ is modelled with a discrete time-to-event model. The additive predictor is defined as
\begin{equation*}
\begin{split}
    \eta = 
    & f_{0}(\text{\code{age}}) + 
    f_{1}(\text{\code{ai}}) + 
    f_{2}(\text{\code{altitude}}) + 
    f_{3}(\text{\code{bord}}) + 
    f_{4}(\text{\code{l\_distance}}) + 
    f_{5}(\text{\code{iyear}}) + \\
    & f_{6}(\text{\code{hhs}}) + 
    f_{7}(\text{\code{higheduyear}}) + 
    f_{8}(\text{\code{magebirth}}) + 
    f_{9}(\text{\code{minc12}}) + 
    f_{10}(\text{\code{ndvi12}}) + \\
    & f_{11}(\text{\code{pre12}}) + 
    f_{12}(\text{\code{l\_rgdp}}) + 
    f_{13}(\text{\code{t2m12}}) + 
    f_{14}(\text{\code{l\_ttcity}}) +
    f_{15}(\text{\code{l\_ttmotor}}) + \\ 
    & f_{16}(\text{\code{l\_ttwalk}}) + 
    f_{17}(\text{\code{lon,lat}}) +
    f_{18}(\text{\code{gender}}) + 
    f_{19}(\text{\code{d\_conf25}}) +
    f_{20}(\text{\code{d\_conf50}}) + \\
    & f_{21}(\text{\code{d\_conf100}}) +
    f_{22}(\text{\code{d\_nl20\_12}}) +
    f_{23}(\text{\code{s\_lc12}}) + 
    f_{24}(\text{\code{s\_soil}}),
\end{split}
\end{equation*}
where $f_{0}, \ldots, f_{24}$ can be a spline, a spatial or a random effect. The variables are described in more detail in the previous section and in Table~\ref{tab:app:vars}.

The model is estimated using BBFIT, following the same specification approach as in the simulation study\:\textendash\:which includes both boosting and resampling steps\:\textendash\:but with an increased number of iterations. Specifically, $L = 1000$ iterations are used in the boosting step to select the most important covariates, and $L = 400$ iterations are used in the resampling step, with the first $200$ discarded as burn-in.

\subsection{Results} \label{sec:app:results}
\begin{figure}[htb!]
\centering
\begin{minipage}{.5\textwidth}
  \centering
  \includegraphics[width=\linewidth]{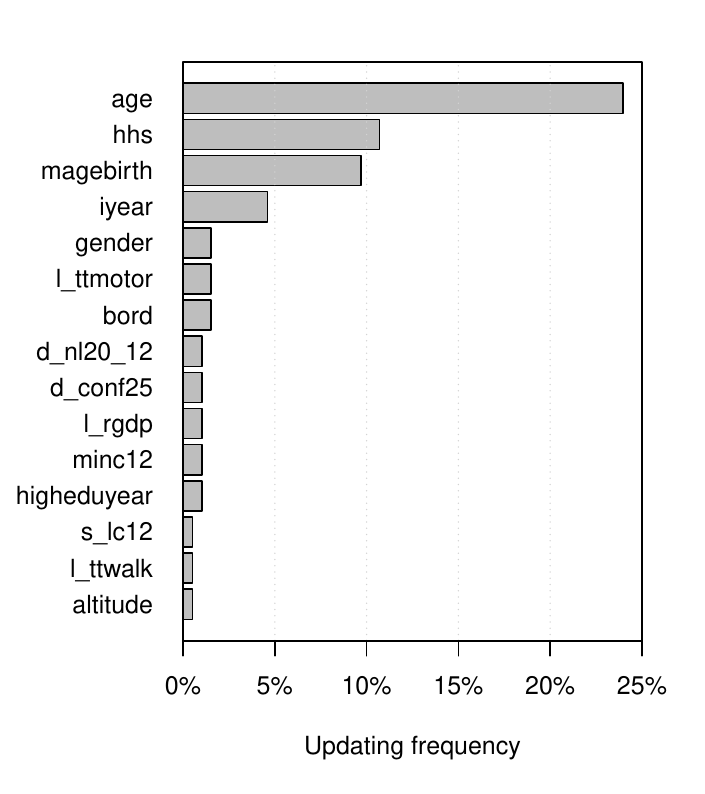}
\end{minipage}%
\begin{minipage}{.5\textwidth}
  \centering
  \includegraphics[width=\linewidth]{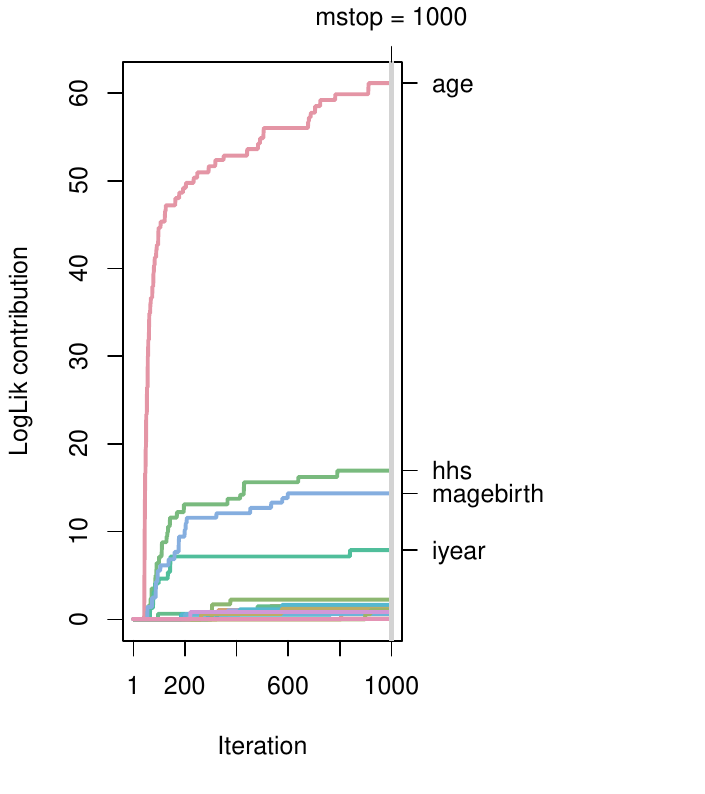}
\end{minipage}
 \vspace{-0.3cm}\caption{\label{fig:app:upfreqscontrib} Updating frequencies based on boosting step of BBFIT (left) and log-likelihood contribution plot (right).  Note that variables that have never been updated are omitted. See Table~\ref{tab:app:vars} for variable description.}
\end{figure}

Figure \ref{fig:app:upfreqscontrib} displays the relative updating frequencies of the boosting step on the left and the log-likelihood contribution on the right. It is the frequency with which a model term yields the best improvement of the out-of-sample log-likelihood by the total number of iterations.

We find that the age of the child (\texttt{age}), the household size (\texttt{hhs}), the age of the mother at birth (\texttt{magebirth}) and the interview year (\texttt{iyear}) are the most important explanatory variables in terms of updating frequency. The age of the child appears to have the by far the greatest importance based on these metrics. While these four variables stand out in terms of importance, the others contribute far less and are not examined further.

\begin{figure}[htb!]
    \centering
    \includegraphics[width=0.48\textwidth]{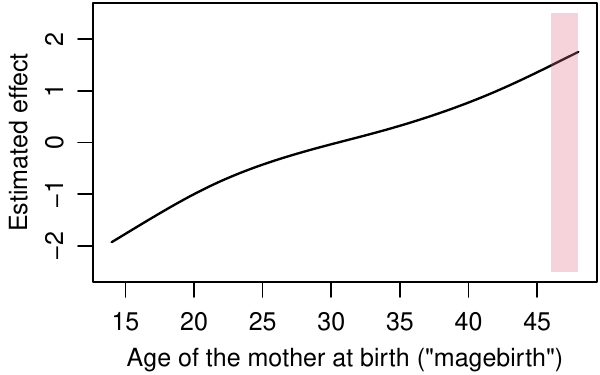}
    \vspace{-0.3cm}\caption{\label{fig:app:magebirtheff} Estimated effects of the covariate \textit{age of the mother at birth} (\texttt{magebirth}) centred at zero. The age groups with fewer than $500$ observations are highlighted in red (see Figure~\ref{fig:app:estivarsmagebirthdes}).}
\end{figure}

As discussed in Section~\ref{sec:app:data}, we focus here first on the covariate \textit{age of the mother at birth} (\texttt{magebirth}) to illustrate the presentation of time-to-event outcomes. Figure~\ref{fig:app:magebirtheff} displays the estimated effect centred at zero.

We find a near-linear increase in mortality risk between ages $18$ and $44$. At the lower and upper ends of the distribution, a slightly steeper increase in risk is visible. In general, this indicates that children with older mothers are more likely to die than children with younger mothers.

\begin{figure}[htb!]
    \centering
    \includegraphics[width=\textwidth]{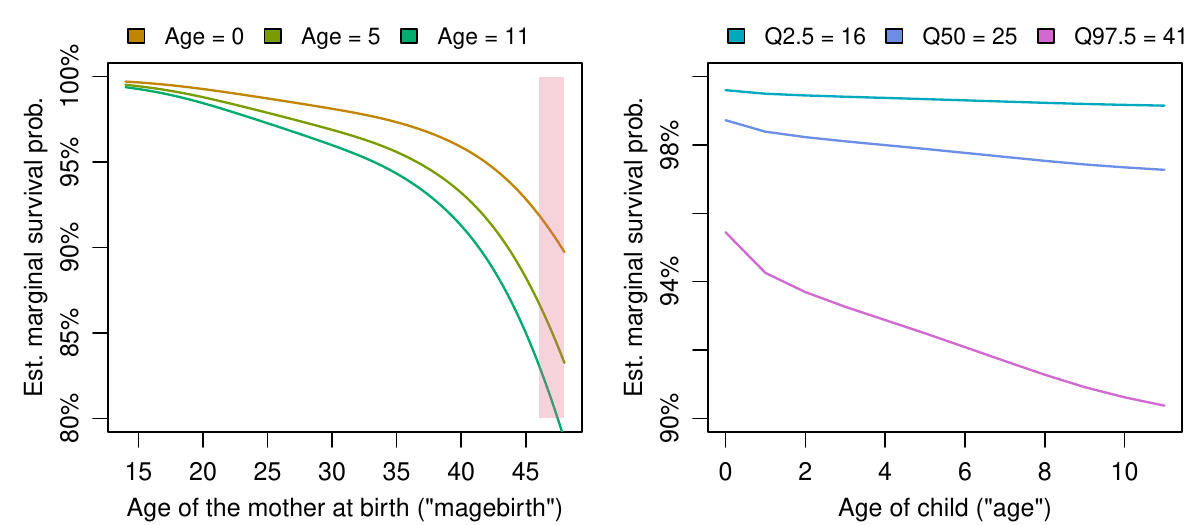}
    \vspace{-0.3cm}\caption{\label{fig:app:magebirthmarsurv} Estimated marginal survival curves for children aged $1$, $5$ and $11$ months, across the age of the mother at birth on the left, and estimated marginal survival probabilities over child age for $16$, $25$ and $41$ ($2.5$th, $50$th, and $97.5$th percentiles respectively) years of maternal age at birth. The age groups with fewer than $500$ observations are highlighted in red (see Figure~\ref{fig:app:estivarsmagebirthdes}).}
\end{figure}

While this trend appears plausible, interpretation on the predictor scale is limited, as it does not directly translate to outcome probabilities and the estimated effects are centred at zero. Of primary interest is how covariates influence the survival probability, as defined in Equation~\eqref{eqn:survival}. Figure~\ref{fig:app:magebirthmarsurv} visualises this relationship for \texttt{magebirth}. All survival probabilities shown are marginal, i.e., all variables, except the one under consideration, are held fixed at their mean (continuous variables) or at the mode (discrete variables) level.

The left panel of Figure~\ref{fig:app:magebirthmarsurv} shows estimated marginal survival curves for children up to an age of one month (Age = 0 corresponding to age interval $[0,1)$), half a year (Age = 5), and one year (Age = 11) corresponding to infant mortality, as a function of the mother’s age at birth. Survival probabilities are close to $100$\% for younger maternal ages but decline notably with increasing maternal age, particularly after age $35$. The downward trend is more pronounced for older children, indicating that the impact of maternal age accumulates over time. 

The right panel shows estimated marginal survival probabilities over child age for three selected maternal ages: $16$ (Q2.5), $25$ (Q50 or median) and $41$ (Q97.5) years. Children born to younger mothers consistently show higher marginal survival probabilities across all ages, while children born to older mothers experience lower survival probabilities, with the gap widening as children age.

These findings underscore the value of variable-specific analyses in understanding complex patterns in survival outcomes. They highlight how flexible modelling approaches benefit from a careful combination of descriptive data exploration (as illustrated in Section~\ref{sec:app:data}) and meaningful interpretation of estimated effects. By examining \texttt{magebirth} in detail, we demonstrate how individual covariates can be linked to outcome dynamics and how model results can be interpreted in a substantively relevant way. Equivalent analyses are provided for all covariates in Appendix \ref{sec:apx:app:singlevar} Figures \ref{fig:apx:app:svhhs} to \ref{fig:apx:app:svaltitude}. Here we continue with a summary of the estimated effects and the estimated marginal probability of survival for the most important variables identified above\:\textendash\:excluding further discussion of \texttt{magebirth}.

\begin{figure}[htb!]
    \centering
    \includegraphics[width=0.8\textwidth]{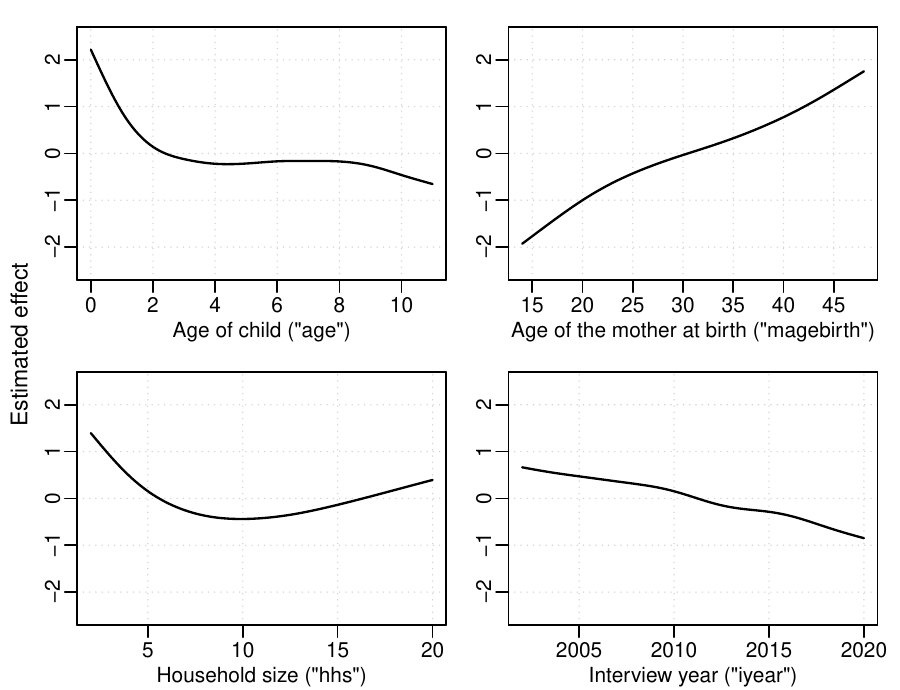}
    \vspace{-0.3cm}\caption{\label{fig:app:estieffnonzero} Estimated effects of the most important variables (see Figure~\ref{fig:app:upfreqscontrib}) centred at zero.}
\end{figure}

Figure \ref{fig:app:estieffnonzero} displays the estimated effects centred at zero. The estimated effect of the age of the child (\texttt{age}) decreases logarithmically over time. This indicates that the probability of dying in the first months of life is greater than later when the first months have been survived. The estimated effect of household size (\texttt{hhs}) follows a U-shape. The probability of dying is lower in households with $5$ to $15$ members and higher in households with fewer or more members. However, the effect is more pronounced in the range $2$ to $4$ household size. One interpretation could be that smaller households go through a learning phase or have limited childcare options due to single parents, while larger households may not have enough attention and resources available for individuals. The estimated effect of the year of interview (\texttt{iyear}) decreases more or less linearly, which is promising as it indicates a decreasing trend in the probability of death over time.

\begin{figure}[htb!]
    \centering
    \includegraphics[width=0.8\textwidth]{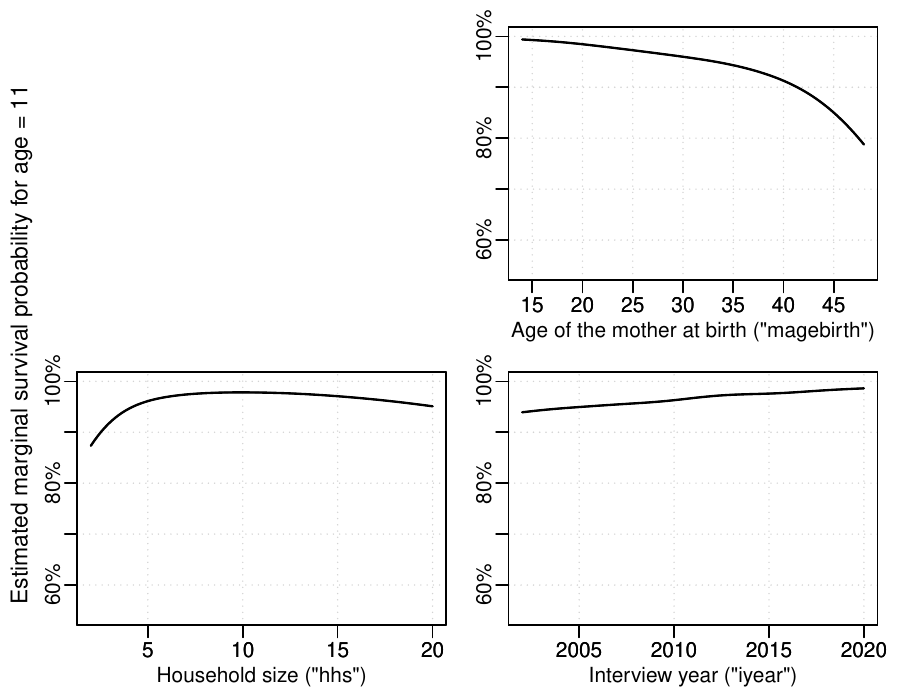}
    \vspace{-0.3cm}\caption{\label{fig:app:estisurvnonzero} Estimated marginal survival probability for the first year of life (Age = 11) of the most important variables (see Figure~\ref{fig:app:upfreqscontrib}).}
\end{figure}

As discussed with the covariate \texttt{magebirth} above, the centred effects are difficult to interpret meaningfully on the predictor scale. Of primary interest is how these factors influence the probability of survival. The estimated marginal probability of survival for the first year of life (Age = 11) is visualised in Figure~\ref{fig:app:estisurvnonzero}.

The three effects (excluding \texttt{age} which is obviously used to construct the survival curves)\:\textendash\:\texttt{magebirth}, \texttt{hhs}, and \texttt{iyear}\:\textendash\:have a substantial influence on the marginal probability of survival. For example, children in small households with less than $5$ members have a much lower marginal probability of surviving the first year of $90\%$ to $95\%$ compared to those in households with $8$ to $10$ members with around $97\%$. For households with more than $15$ members, the marginal probability of surviving the first year decreases again to around $95\%$. Similar interpretations are possible for the other explanatory variables. The estimated effects and the marginal survival curves for all selected effects are shown in Appendix \ref{sec:apx:app:estisurv} in Figure~\ref{fig:apx:appesti} and \ref{fig:apx:appsurv}.

% \clearpage
% \newpage
% conclusion --------------------
\section{Conclusion}\label{sec:conclusion}
This paper introduces an efficient estimation framework for discrete time-to-event models with additive predictors by extending the recently proposed Batchwise Backfitting algorithm (BBFIT). Our contribution lies in combining scalable estimation with simultaneous variable selection in high-dimensional survival settings\:\textendash\:addressing key limitations of existing approaches.

Through a comprehensive simulation study, we demonstrate that the proposed method (BBFIT) achieves high estimation accuracy, excellent variable selection performance, and significantly reduced computation times, even with data sets comprising up to ten million rows. Compared to benchmark methods such as GLM and BAM, BBFIT shows superior robustness and efficiency, especially in high-dimensional or large-scale contexts. In particular, BBFIT consistently avoids the selection of non-informative covariates, while BAM and GLM have a non-negligible false selection rate.

The application to modelling infant mortality in ten countries in eastern sub-Saharan Africa further illustrates the practical utility of our approach. Despite the complexity of the data\:\textendash\:both in terms of sample size and number of covariates\:\textendash\:BBFIT identifies a small subset of influential factors (e.g., child's age, household size, and survey year) and provides interpretable smooth effect estimates. This real-world example highlights how BBFIT can enhance predictive performance and yield meaningful insights in public health and development research, while also demonstrating its applicability to a broad range of real-world applications.

While the proposed framework is a powerful tool for large-scale discrete time-to-event modelling, several directions for future work remain. For instance, exploring alternative model structures (e.g., piecewise exponential models or Box-Cox-transformed hazard models) could expand its applicability beyond the current link and distributional assumptions. Another potential extension is the development of principled importance measures to quantify the relevance of selected predictors beyond selection frequencies or updating rates. 

On the computational side, future work may focus on extending the algorithm towards a Bayesian framework to account for the model uncertainty. Furthermore, exploiting the specific structure of the likelihood\:\textendash\:where all entries are zero except possibly the last\:\textendash\:could lead to additional algorithmic efficiencies. These enhancements would enable more informative inference and further improve the scalability of the method for application to more complex settings in modern time-to-event analysis.

% coi --------------------
\section*{Disclosure statement}
No potential conflict of interest was reported by the authors.

% acknowledgements ------------------
\section*{Acknowledgements} The computational results presented have been achieved (in part) using the HPC infrastructure LEO of the University of Innsbruck. We thank ICF International, Inc. and USAID for conducting the DHS \citep{ICF:2004-2017}, and providing public access to the data. This research was partially funded by the Austrian Science Fund (FWF) grant. doi: \href{https://doi.org/10.55776/P33941}{10.55776/P33941}.

% software ------------------
\section*{Software} The computational results presented have been derived using the statistical software \proglang{R} \citep{RCore2020} using the following attached \proglang{R} packages: \pkg{bamlss} \citep{umlauf_bamlss_2018, umlauf_bamlss_2021, umlauf_scalable_2024}, \pkg{mgcv} \citep{wood_thin_2003, wood_fast_2011, wood_smoothing_2016, wood_generalized_2017-1}, \pkg{MASS} \citep{venables_modern_2002}, \pkg{discSurv} \citep{welchowski_discsurv_2022}, \pkg{sf} \citep{pebesma_simple_2018, pebesma_spatial_2023}, \pkg{colorspace} \citep{stauffer_somewhere_2009, zeileis_escaping_2009, zeileis_colorspace_2020}, \pkg{lubridate} \citep{grolemund_dates_2011}.

\clearpage
\pagebreak

% bibliography ------------------
%% references need to be provided in a .bib BibTeX database
%% all references should be made with \cite, \citet, \citep, \citealp etc.
%%  (and never hard-coded). See the FAQ for details
%% jss-specific markup (\proglang, \pkg, \code) should be used in the .bib
%% titles in the .bib should be in title case
%% dois should be included where available
% \bibliographystyle{chicago} already set in jss.cls
\bibliography{bb4sa}

\clearpage
\pagebreak

% appendix (if any) ------------------
%% after the bibliography with page break
%% with proper section titles and _not_ just "appendix"
\begin{appendix}
%\counterwithin{figure}{section}

\section{Appendix: Simulation results} \label{sec:appendixsim}
\subsection{Basic setting: Noise/uninformative variables} \label{sec:apx:sim:basic}

\begin{figure}[htb!]
    \centering
    \includegraphics[width=0.85\textwidth]{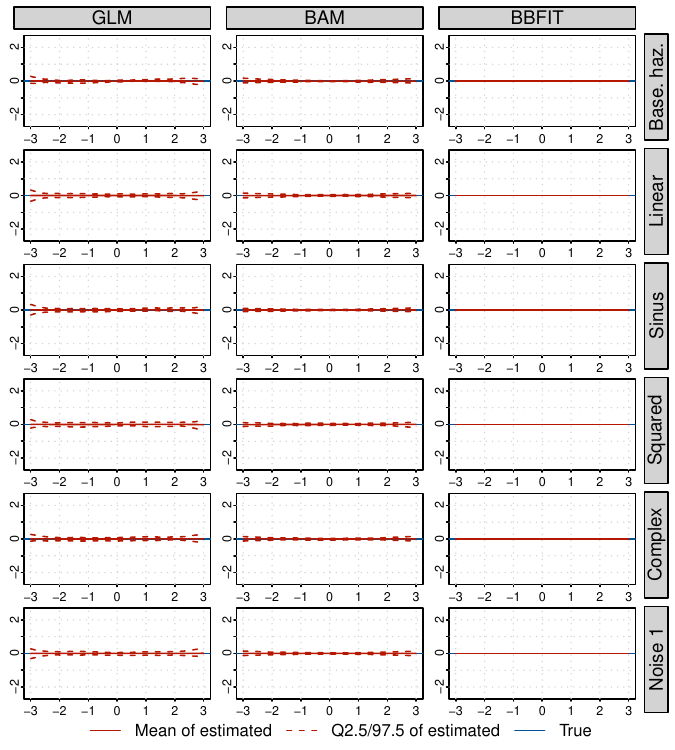}
    \vspace{-0.3cm}\caption{\label{fig:apx:basicestinoise} Estimated effects of all noise/uninformative variables in the basic setting (see Section \ref{sec:sim:designbasic}).}
\end{figure}

\begin{figure}[htb!]
    \centering
    \includegraphics[width=0.85\textwidth]{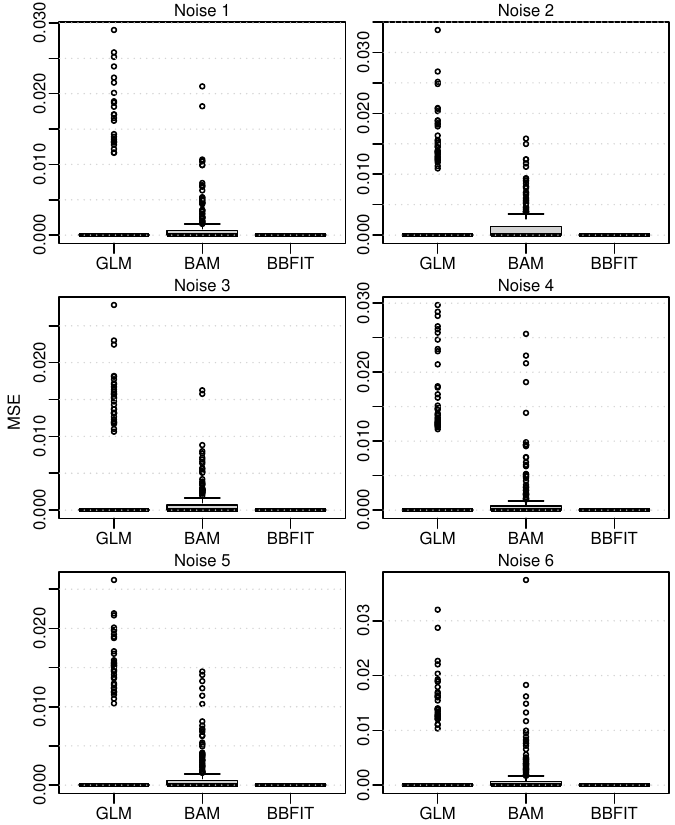}
    \vspace{-0.3cm}\caption{\label{fig:apx:basicrmsenoise} MSE of all noise/uninformative variables in the basic setting (see Section \ref{sec:sim:designbasic}).}
\end{figure}

\clearpage
\pagebreak

\subsection{Further settings: Mean MSE by number of individuals} \label{sec:apx:sim:furth}

\begin{figure}[htb!]
    \centering
    \includegraphics[width=\textwidth]{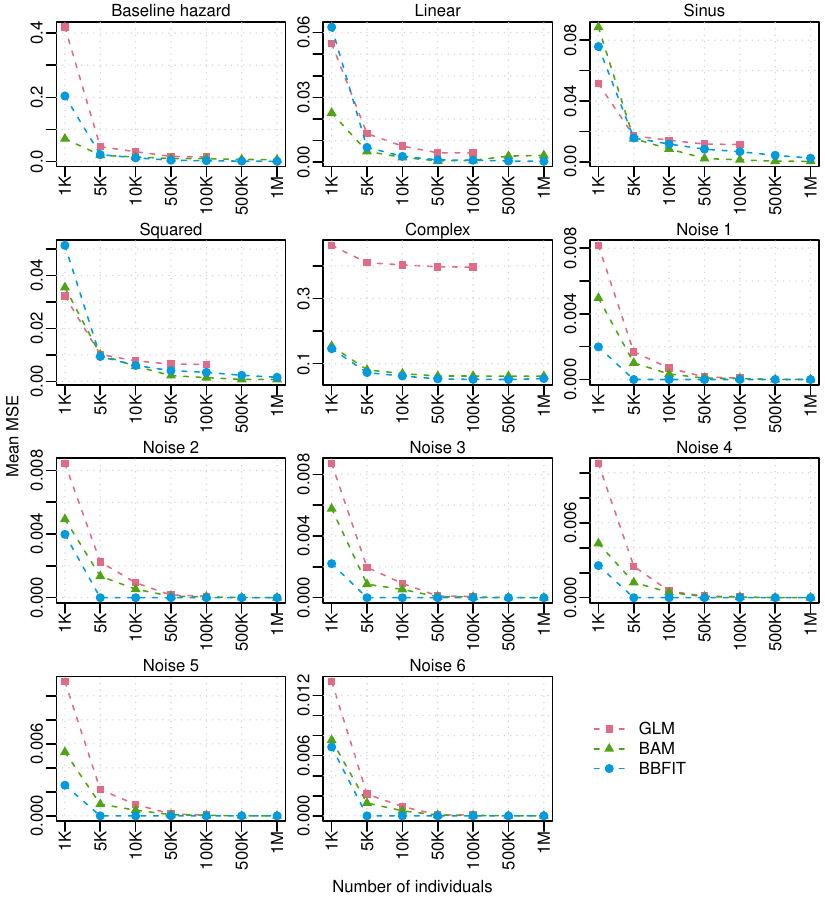}
    \vspace{-0.3cm}\caption{\label{fig:apx:furthrmseall} Mean MSE for different number of individuals of all variables.}
\end{figure}

\clearpage
\pagebreak

\section{Appendix: Application results} \label{sec:appendixapp}
\subsection{Single variable analysis} \label{sec:apx:app:singlevar}

\begin{figure}[htb!]
    \centering
    \includegraphics[width=\textwidth]{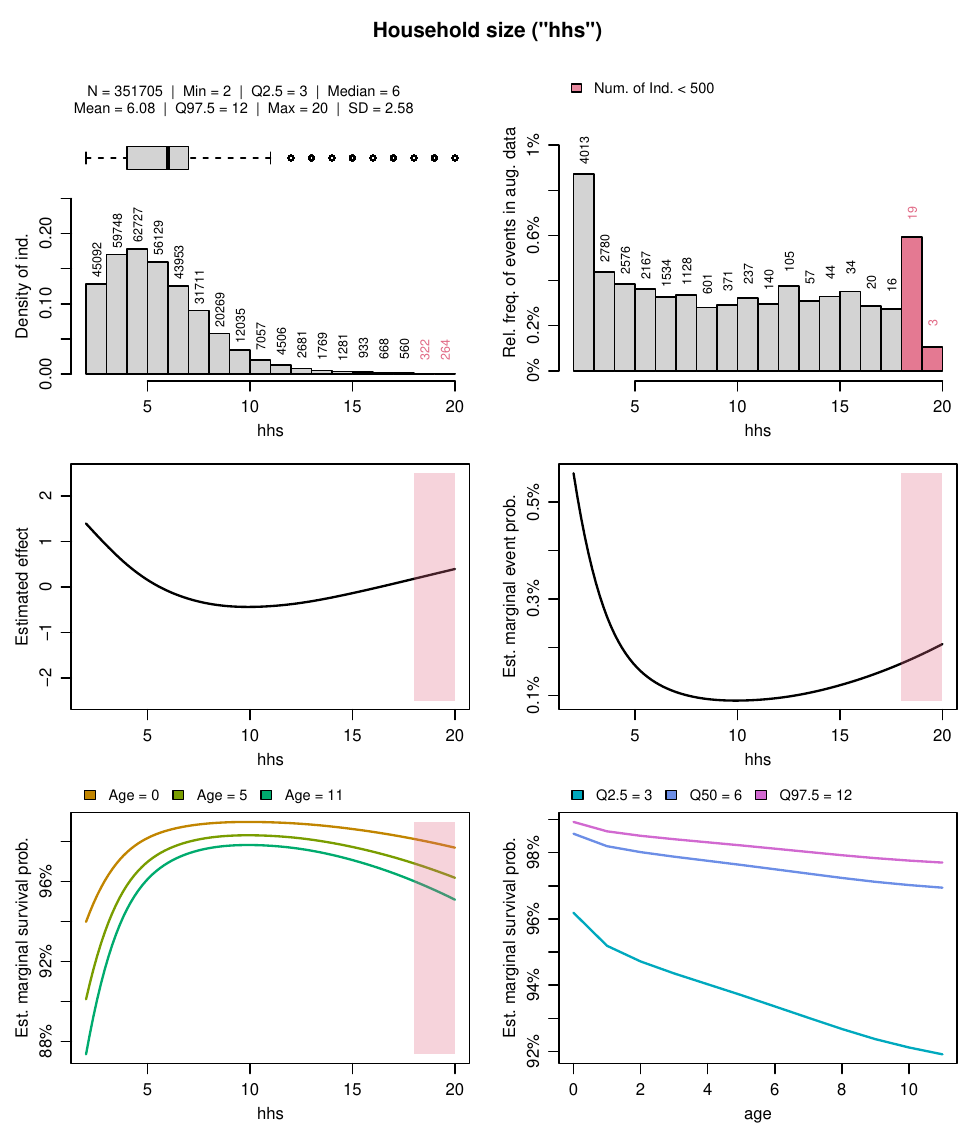}
    \vspace{-0.3cm}\caption{\label{fig:apx:app:svhhs} Single variable analysis of covariate \textit{household size} (\texttt{hhs}).}
\end{figure}

\clearpage
\pagebreak

\begin{figure}[htb!]
    \centering
    \includegraphics[width=\textwidth]{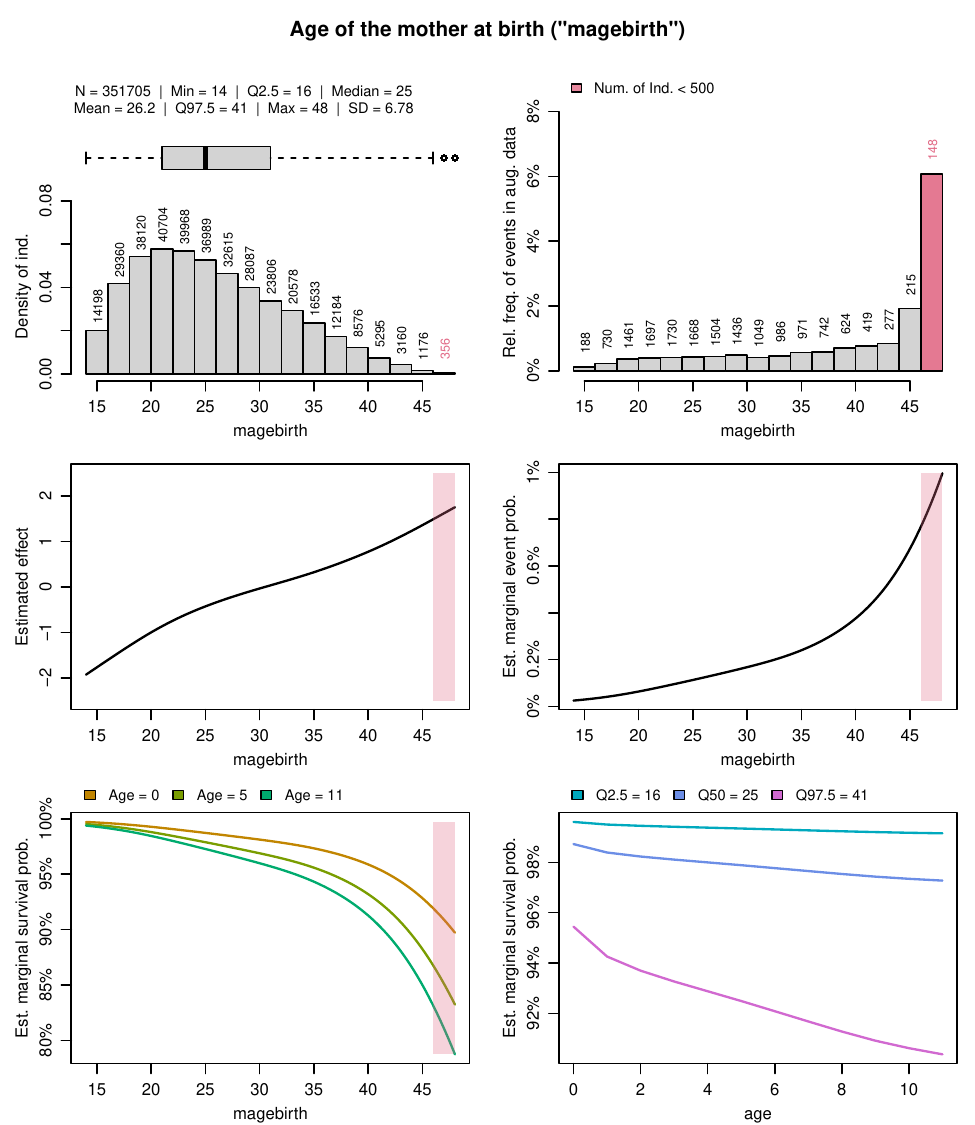}
    \vspace{-0.3cm}\caption{\label{fig:apx:app:svmagebirth} Single variable analysis of covariate \textit{age of the mother at birth} (\texttt{magebirth}).}
\end{figure}

\clearpage
\pagebreak

\begin{figure}[htb!]
    \centering
    \includegraphics[width=\textwidth]{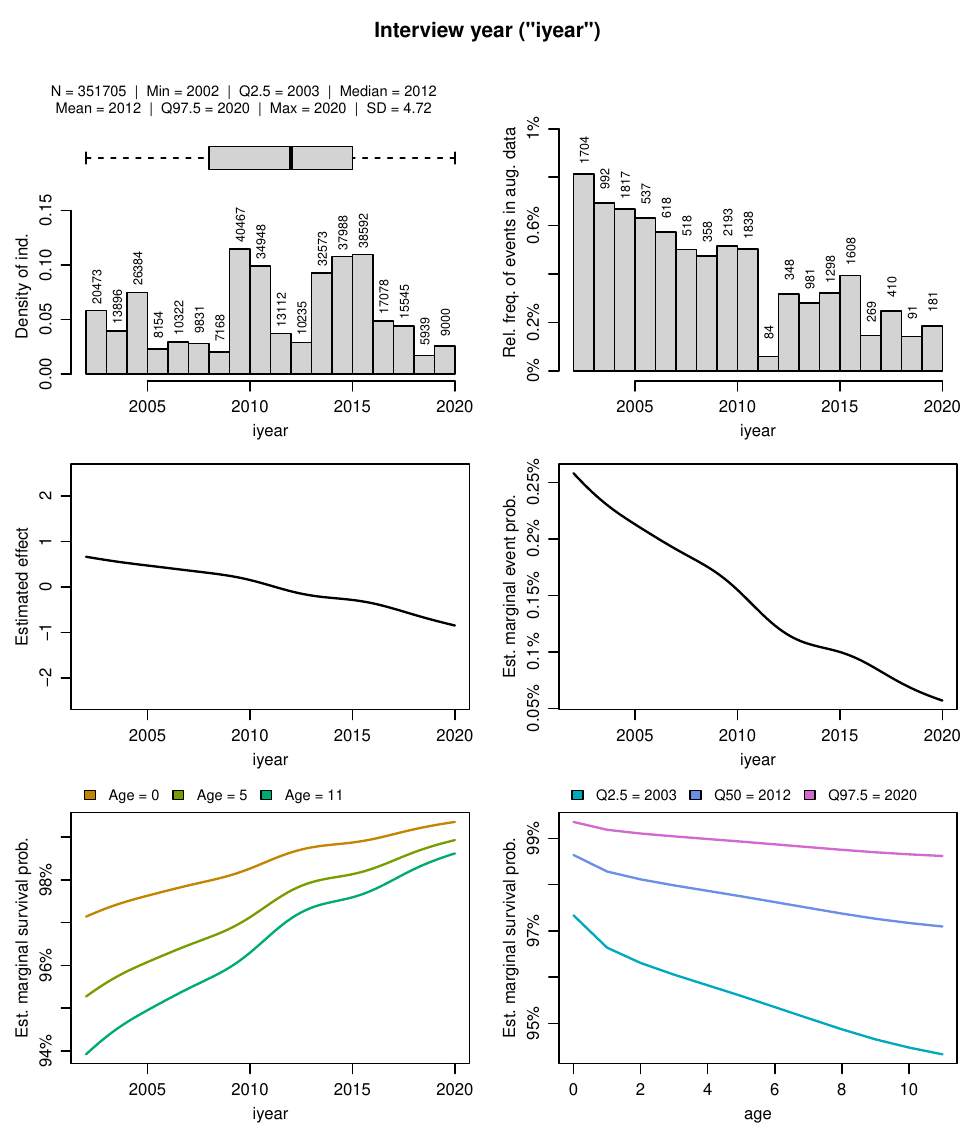}
    \vspace{-0.3cm}\caption{\label{fig:apx:app:sviyear} Single variable analysis of covariate \textit{interview year} (\texttt{iyear}).}
\end{figure}

\clearpage
\pagebreak

\begin{figure}[htb!]
    \centering
    \includegraphics[width=\textwidth]{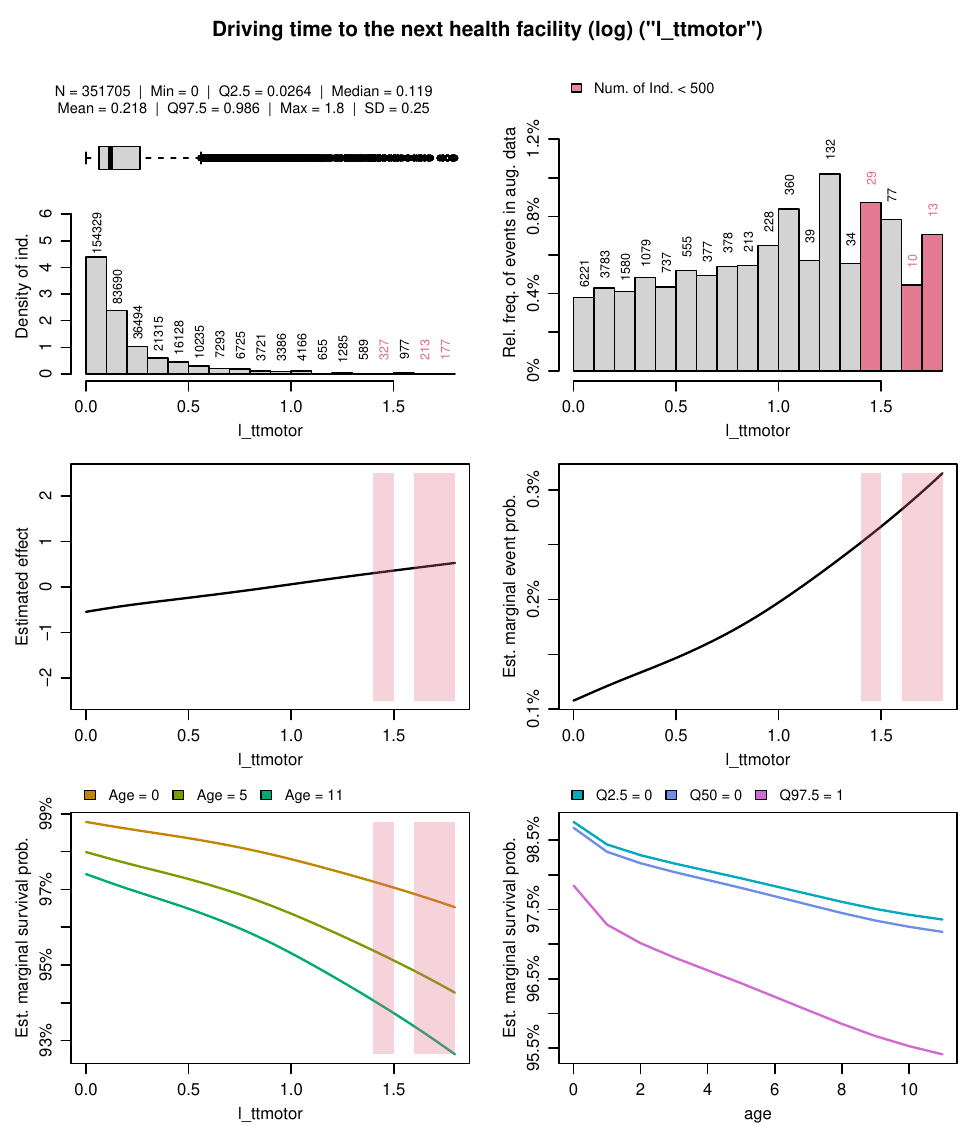}
    \vspace{-0.3cm}\caption{\label{fig:apx:app:svl_ttmotor} Single variable analysis of covariate \textit{driving time to the next health facility (log)} (\texttt{l\_ttmotor}).}
\end{figure}

\clearpage
\pagebreak

\begin{figure}[htb!]
    \centering
    \includegraphics[width=\textwidth]{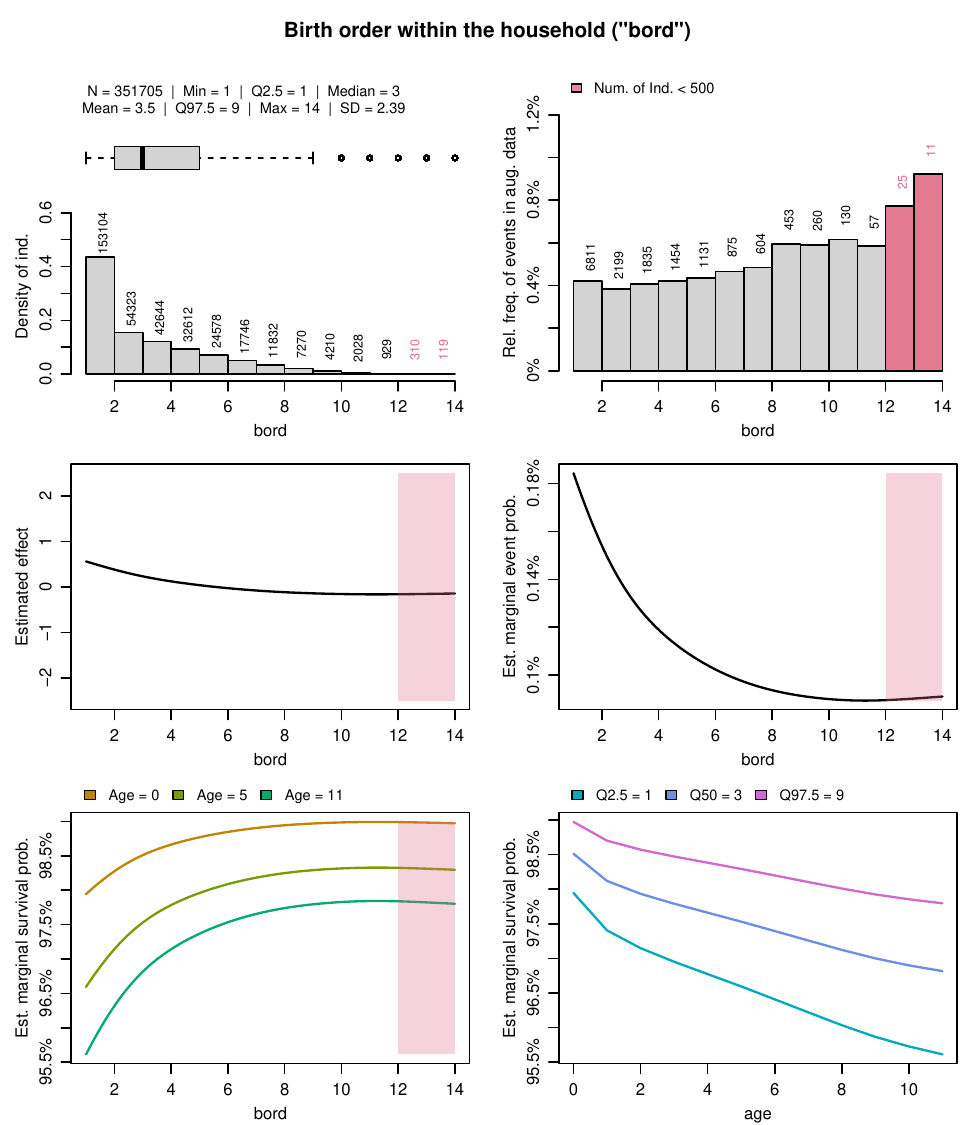}
    \vspace{-0.3cm}\caption{\label{fig:apx:app:svbord} Single variable analysis of covariate \textit{birth order within the household} (\texttt{bord}).}
\end{figure}

\clearpage
\pagebreak

\begin{figure}[htb!]
    \centering
    \includegraphics[width=\textwidth]{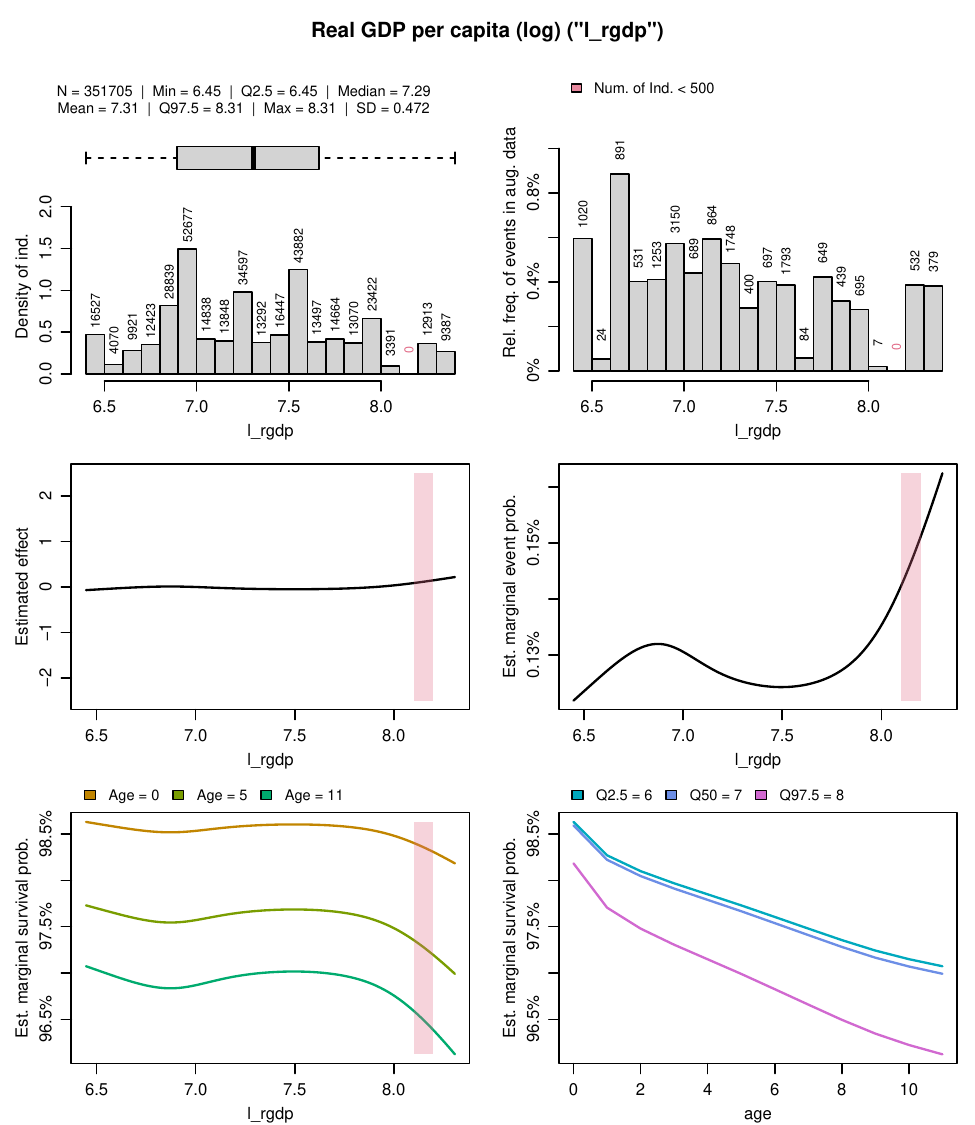}
    \vspace{-0.3cm}\caption{\label{fig:apx:app:svl_rgdp} Single variable analysis of covariate \textit{real GDP per capita (log)} (\texttt{l\_rgdp}).}
\end{figure}

\clearpage
\pagebreak

\begin{figure}[htb!]
    \centering
    \includegraphics[width=\textwidth]{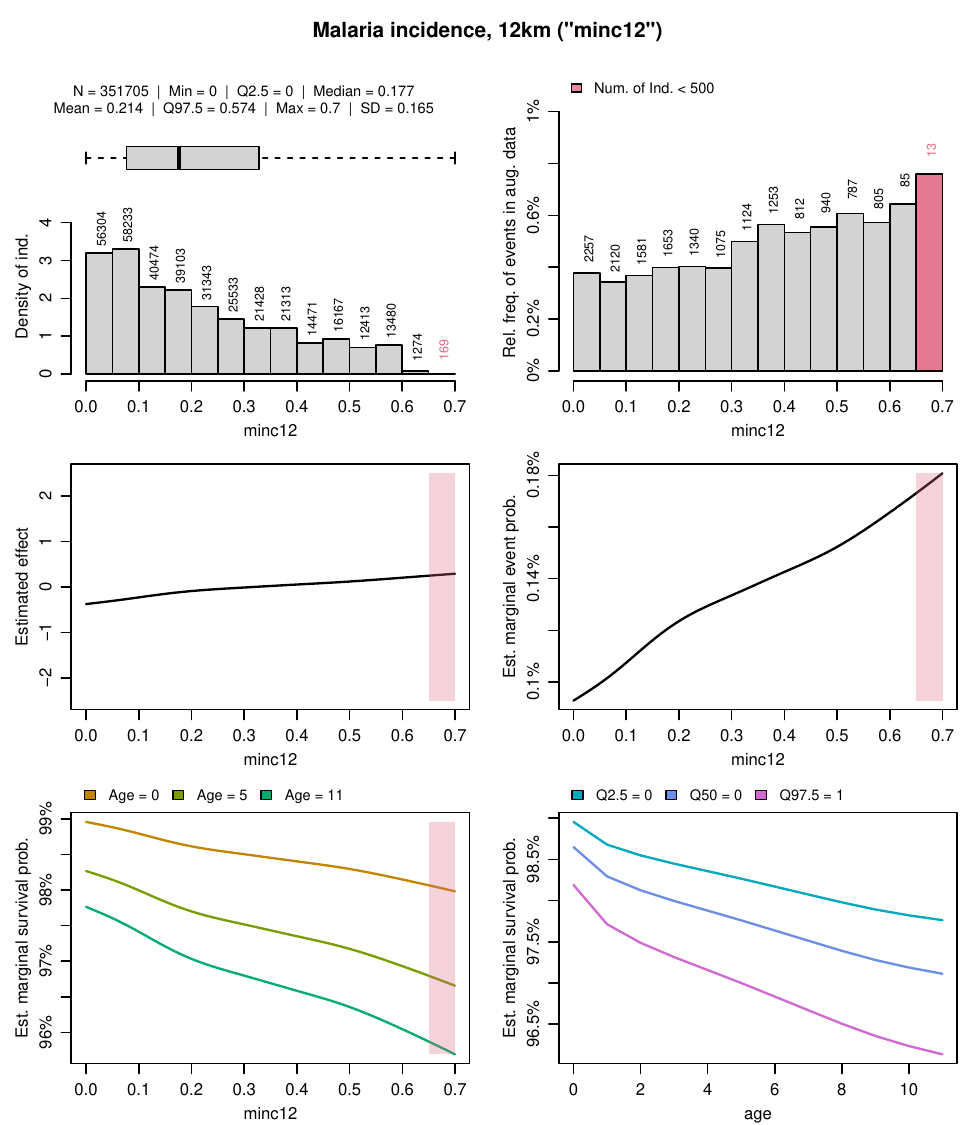}
    \vspace{-0.3cm}\caption{\label{fig:apx:app:svminc12} Single variable analysis of covariate \textit{malaria incidence, 12km} (\texttt{minc12}).}
\end{figure}

\clearpage
\pagebreak

\begin{figure}[htb!]
    \centering
    \includegraphics[width=\textwidth]{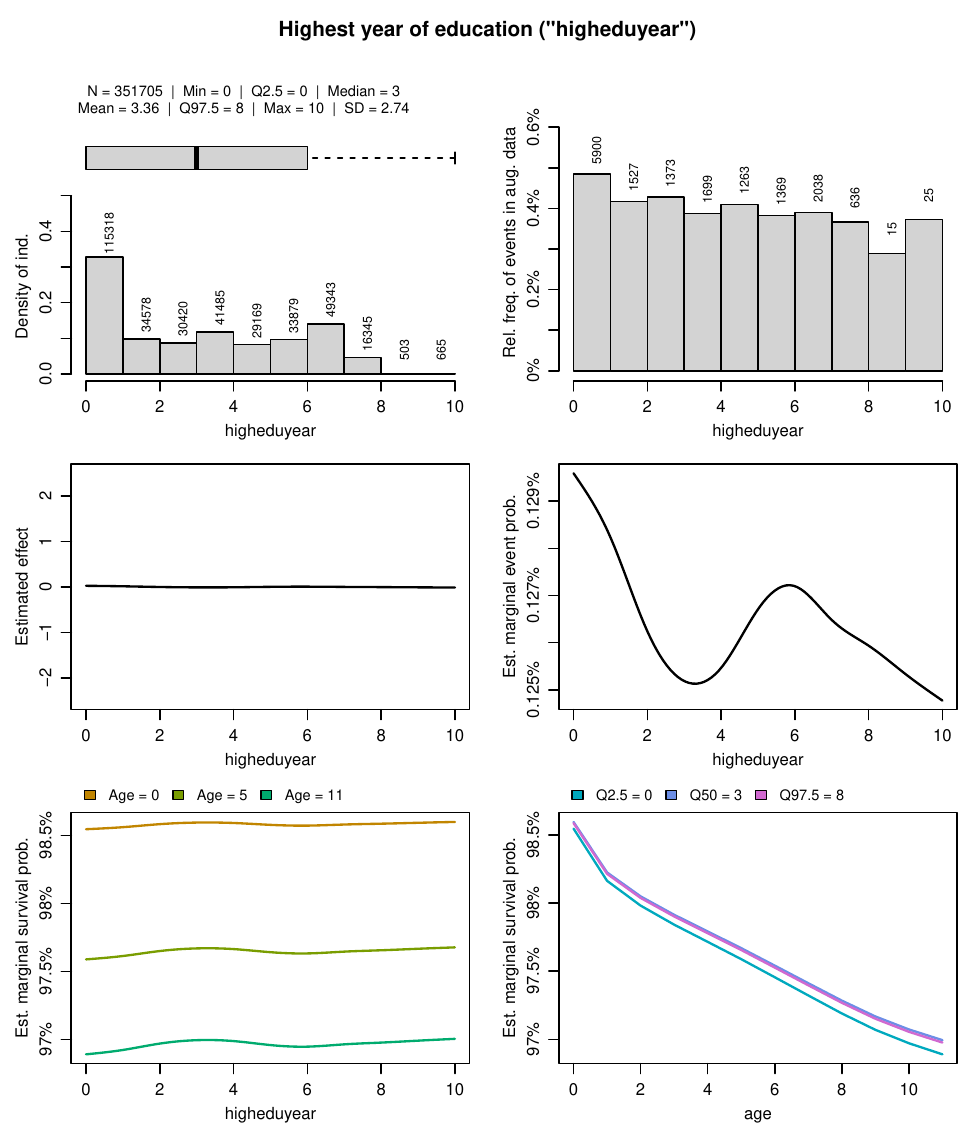}
    \vspace{-0.3cm}\caption{\label{fig:apx:app:svhigheduyear} Single variable analysis of covariate \textit{highest year of education} (\texttt{higheduyear}).}
\end{figure}

\clearpage
\pagebreak

\begin{figure}[htb!]
    \centering
    \includegraphics[width=\textwidth]{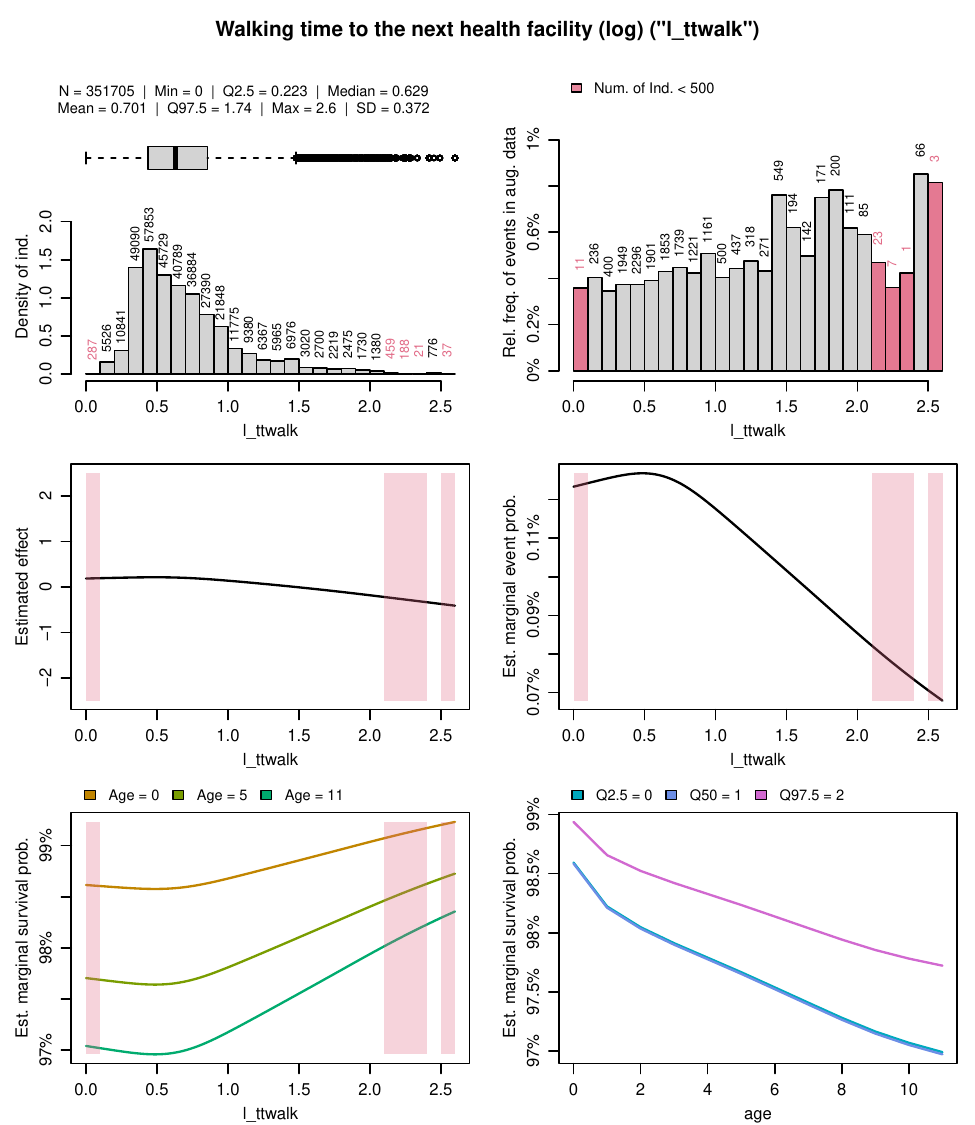}
    \vspace{-0.3cm}\caption{\label{fig:apx:app:svl_ttwalk} Single variable analysis of covariate \textit{walking time to the next health facility (log)} (\texttt{l\_ttwalk}).}
\end{figure}

\clearpage
\pagebreak

\begin{figure}[htb!]
    \centering
    \includegraphics[width=\textwidth]{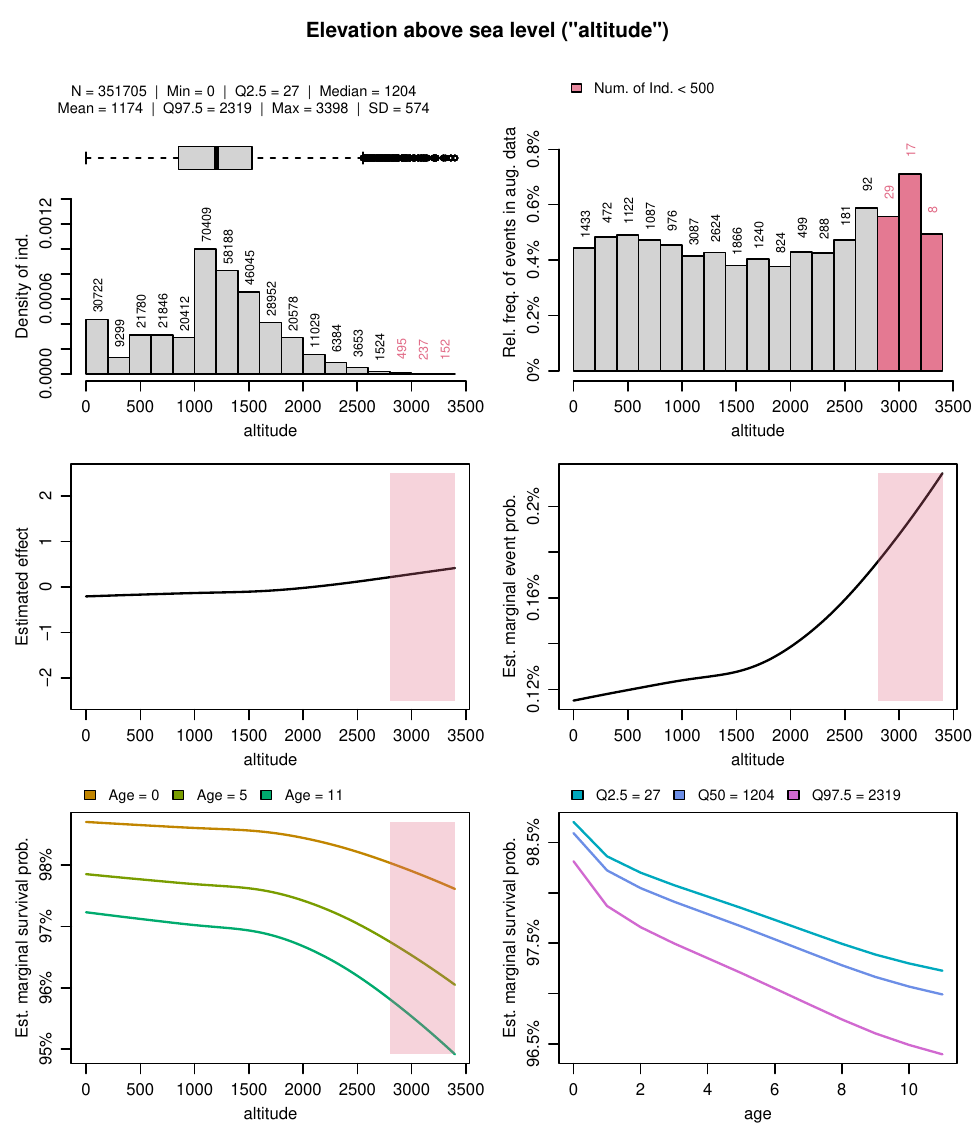}
    \vspace{-0.3cm}\caption{\label{fig:apx:app:svaltitude} Single variable analysis of covariate \textit{elevation above sea level} (\texttt{altitude}).}
\end{figure}

\clearpage
\pagebreak

\subsection{Estimated effects and marginal probability of survial} \label{sec:apx:app:estisurv}

\begin{figure}[htb!]
    \centering
    \includegraphics[width=\textwidth]{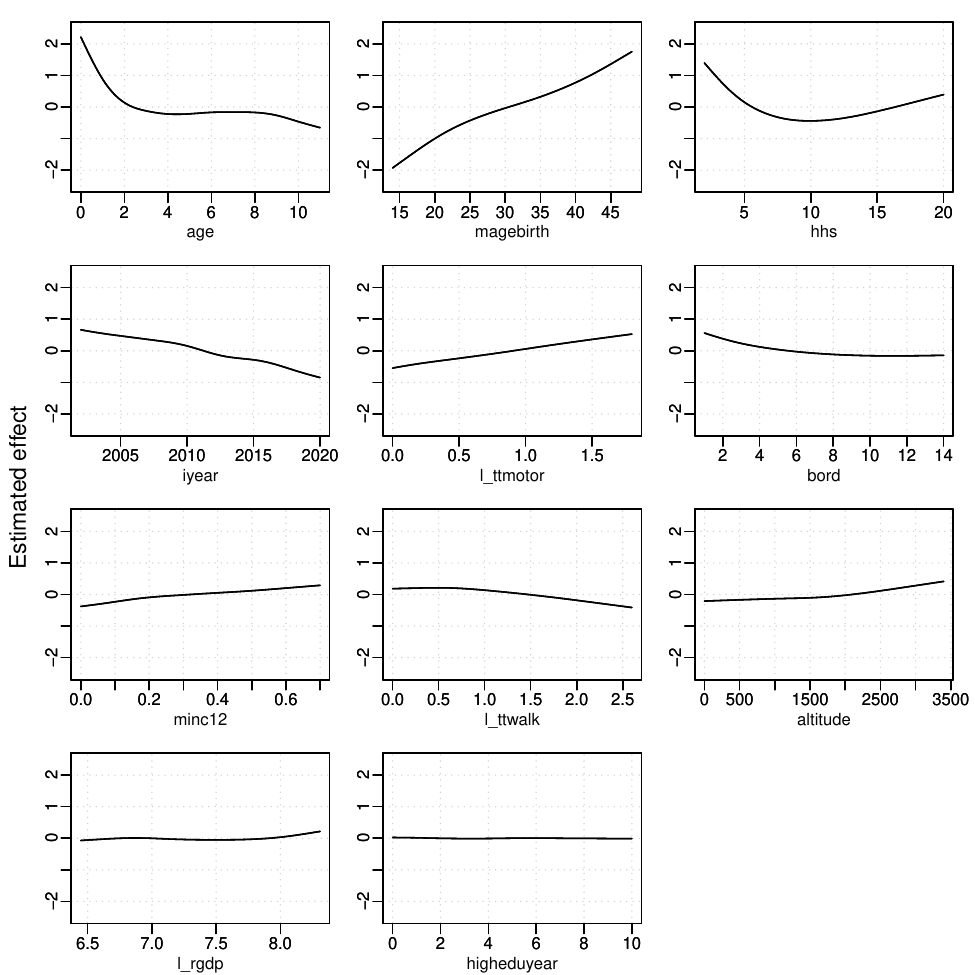}
    \vspace{-0.3cm}\caption{\label{fig:apx:appesti} Estimated effects of all selected variables (see Figure~\ref{fig:app:upfreqscontrib}) centred at zero. See Table~\ref{tab:app:vars} for variable describtion.}
\end{figure}

\begin{figure}[htb!]
    \centering
    \includegraphics[width=\textwidth]{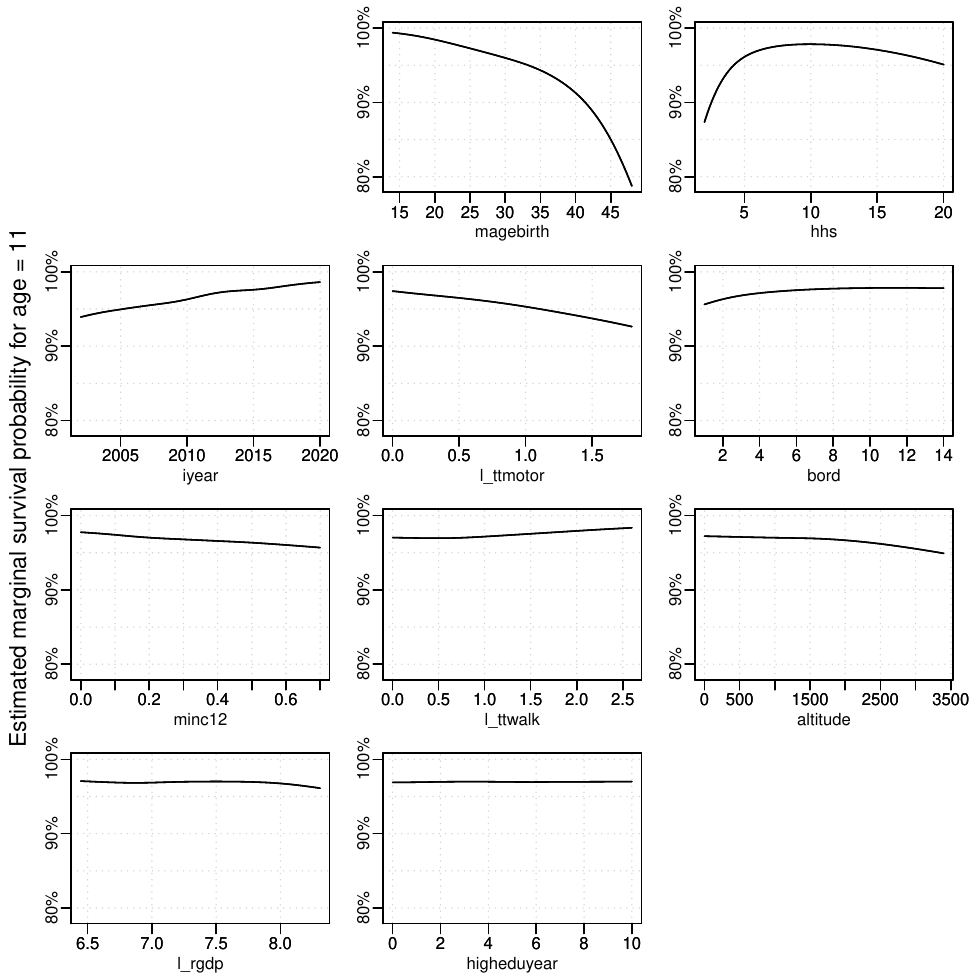}
    \vspace{-0.3cm}\caption{\label{fig:apx:appsurv} Estimated marginal survival probability for the first year of life (Age = 11) of all selected variables (see Figure~\ref{fig:app:upfreqscontrib}). See Table~\ref{tab:app:vars} for variable describtion.}
\end{figure}

\clearpage
\pagebreak

\end{appendix}
\end{document}